\documentclass[aps,preprint]{revtex4-1}

\usepackage[dvipdf,]{graphicx}
\usepackage{graphicx}
\usepackage{float}
\usepackage{color}
\usepackage[caption=false]{subfig}
\usepackage{amsmath}
\usepackage{siunitx}
\usepackage{hyperref}

\newcommand{\md}{\mathrm{d}}
\newcommand{\me}{\mathrm{e}}
\newcommand{\mi}{\mathrm{i}}

\renewcommand{\tt}[1]{\mathrm{#1}}
\renewcommand{\vec}[1]{\mathbf{#1}}

\newcommand{\eq}{\begin{equation}}
\newcommand{\qe}{\end{equation}}

\renewcommand{\k}{\vec{k}}
\newcommand{\q}{\vec{q}}
\newcommand{\Q}{\vec{Q}}
\newcommand{\G}{\vec{G}}
\newcommand{\rr}{\vec{r}}

\newlength{\figwidth}
\newlength{\widefigwidth}
\begin{document}
\title{Validating x-ray line-profile defect analysis using atomistic models of deformed material}
\author{C P Race}
\email[Contact: ]{christopher.race@manchester.ac.uk}
\author{T Ungar}
\affiliation{Department of Materials, University of Manchester, Manchester, M13 9PL, United Kingdom}
\author{G Rib\'arik}
\affiliation{Department of Materials Physics, E\"otv\"os Lor\'and University Budapest, H-1117 P\'azm\'any P. s\'et\'any 1/A, Hungary}

\begin{abstract}
The population of dislocation defects in a crystalline material strongly influences its properties, so the ability to analyse this population in experimental samples is of great utility. As a complement to direct counting in the transmission electron microscope, quantitative analysis of x-ray diffraction line profiles is an important tool. This is an indirect approach to quantification and so requires careful validation of the physical models that underly the inferential process. Here we undertake to directly evaluate the ability of line profile analysis to quantify aspects of the dislocation and stacking fault populations by exploiting atomistic models of deformed copper single crystals. We directly analyse these models to determine exact details of the defect content (our ``ground truth''). We then generate theoretical line profiles for the models and analyse them using the same procedures used in experimental analysis. This leads to inferred measures of the defect content which we are able to compare with the exact data. We show that line profile analysis is able to provide sound predictions of both dislocation density and stacking fault fraction across two orders of magnitude. We further show how the outer cut-off radius in the mean-square strain of a dislocation distribution invoked by Warren and Averbach corresponds to the cell size in an artificially constructed restrictedly-random distribution of dislocations according to the model of Wilkens. Overall, our results lend important new support to the use of line profile analysis for the quantification of line and planar defects in crystalline materials.
\end{abstract}

\maketitle

%\tableofcontents

\section{Introduction}\label{sec:intro}

Dislocation defects are, of course, a hugely important feature of polycrystalline materials: via their motion they are the mediators of plastic deformation; their presence can change the basic properties of the material (e.g.\ via work-hardening); by raising the free energy of deformed material they provide a driving force for recrystallisation; they can provide a mechanism of failure (e.g.\ as when a dislocation pile-up gives rise to crack formation); and, they constitute `pipes' for the more rapid distribution of solutes around a microstructure.

Hence a detailed understanding of the formation and evolution of populations of dislocations in poly-crystalline materials is needed when seeking to control their behaviour in processing and in service. This means that we need a way to `measure' dislocations: an experimental means of determining what types of dislocations are present, how many of them there are and whereabouts in the microstructure they lie. 

The most direct way of quantitatively analysing dislocations is to image them in a transmission electron microscope, and make measurements of the resulting micrographs, varying the diffraction conditions in order to determine the dislocation character through changes in contrast. This is a highly successful approach, but nevertheless has several potential drawbacks:
\begin{enumerate}
\item the process of preparing foils is time consuming and can be expensive (e.g.\ for active materials);
\item foil preparation can do considerable violence to the material, leading to changes in the dislocation distribution that we are seeking to measure, not least because the large amount of surface in a thin foil is an efficient sink for dislocation annihilation;
\item some dislocations that are important for the processes under investigation may be too small to resolve in the TEM;
\item beyond a certain density ($\rho\sim 10^{14}\,\tt{m}^{-2}$~\cite{Wilkens:1970aa}) individual dislocations become difficult to resolve and the presence of other defects, in the bulk and on the surface of the foil, can exacerbate this issue;
\item the total volume of material analysed in a TEM is typically rather small (a good usable area of $1\,\mu\textrm{m}^2$ in a thickness of $100\,\textrm{nm}$ would entail 10 billion foils for a volume of only $1\,\textrm{mm}^3$) leading to sometimes poor counting statistics;
\item counting and measurement of the dislocations is usually a manual process, introducing subjectivity into the results.
\end{enumerate}

An alternative way of quantifying dislocation content is provided by the practice of x-ray \emph{line profile analysis} (LPA). In this approach the one-dimensional profile of the scattered intensity as a function of scattering angle is produced, as an integral around the Debye-Scherrer rings formed by diffraction of a highly mono-chromated beam of x-rays by the sample under study. Such profiles are most often derived from powder or polycrystalline samples, but with a sufficiently bright source, such as that provided by a synchrotron, the profiles for individual crystallites (grains) can also be produced. X-ray line-profiles are commonly used in determining crystal structures and macro-strains by analysing the \emph{positions} of the Bragg peaks in the profile, each corresponding to the \emph{spacing} of a particular set of crystallographic planes in the sample. LPA extends this approach by analysing the \emph{shapes} of these peaks, which are determined by the \emph{distortion} of those planes and so to the types of defect present in the material and causing non-uniform atomic displacement.

Though the LPA approach addresses, at least in part, all the issues with direct defect counting in the TEM, it has deficiencies of its own and should be regarded as a complementary technique, rather than a replacement. The most significant issue is that LPA is an \emph{indirect} method of determining dislocation content. We will discuss the theory in more detail in Section~\ref{sec:methods:lpa}, but broadly speaking, LPA proceeds by using a physical model of the atomic displacement field caused by a particular defect type and calculating the effect that the corresponding lattice strain will have on the shapes of the Bragg peaks. We thus face an \emph{inverse problem} of taking the peak shapes and attempting to infer the nature, quantity and arrangement of the defects that cause them. This would be a hopeless task if it were not for the fact that the response of the peak shape varies from peak to peak, with diffraction order and with the distance from the peak centre, in different ways as we consider dislocations with different burgers vectors, characters (screw versus edge) and arrangements. Nevertheless, the correspondence is neither simple nor one-to-one and we invariably need to know at least a little about the defects present before attempting a line profile analysis (hence its being a complement to TEM observations).

A second issue is the nature of the physical models describing the lattice strain due to dislocations and the effect of a distribution of many dislocations on the peak shape. The distortion due to an \emph{individual} dislocation is well described by theory at various levels of physics, including for the case of a general burgers vector and line direction, and for anisotropic elasticity \cite{Bacon:1980fk}. But when many dislocations are present, the peak shape is rather sensitive to how those dislocations are \emph{arranged}. Again, we will consider this in more detail in Section~\ref{sec:methods:lpa}, but for now we remark that the effect of dislocation arrangement is generally captured through an upper bound, $R_{\tt{e}}$, on the range of the strain field for each individual dislocation, related to the stored energy of dislocations. This has variously been interpreted as corresponding to the grain size \cite{Krivoglaz:1963aa}, the characteristic spacing between dislocations \cite{Wilkens:1968aa} or, in the case of Wilkens' \emph{restrictedly-random distribution}~\cite{Wilkens:1969ab,Wilkens:1969ac}, to a length-scale associated with details of the dislocation distribution (via the $M$ parameter, to be defined below). Unfortunately, the physical meaning of $R_{\tt{e}}$ (or, equivalently, of $M$) is clear only in a few highly idealised cases and so it is generally treated as an independent fitting parameter in the LPA process. A more general interpretation is a long-standing problem.

Despite the solid grounding in physical models of lattice distortion, the inverse nature of LPA and the residual uncertainties in the meaning of $R_{\tt{e}}$ mean that validation of inferred properties of the dislocation distributions is required. In the past, this has most often been performed via comparison with direct counts from the TEM. For example, dislocation densities in copper single crystals oriented for single slip and tensile deformed to different strain levels up to about 70 MPa have been determined by TEM \cite{Essmann:1966vd,Essmann:1965wi,Gottler:1973uh} and X-ray line broadening \cite{Ungar:1984aa,Wilkens:1968aa}. X-ray line broadening was evaluated using the Krivoglaz-Wilkens approach \cite{Krivoglaz:1963aa,Ungar:1984aa,Wilkens:1968aa,Wilkens:1969ab,Wilkens:1969ac} and the dislocation density values obtained by the two different methods were identical within experimental error. In Fig.\ 2 of Ref.\ \cite{Wilkens:1968aa} it was shown that in the low density range TEM gives reliable dislocation densities \cite{Essmann:1966vd,Essmann:1965wi}, however, when the local dislocation density exceeds about $10^{14}-10^{15}\,\textrm{m}^{-2}$ TEM counting becomes problematic while X-ray line broadening can still provide reliable values. Such comparisons have successfully established LPA as a valuable tool for studying defect populations. But as discussed above, TEM analysis itself is subject to limitations and a second, independent means of validating LPA would be a valuable addition.

Continual increases in computing power mean that atomistic simulation can now provide a direct means of validating LPA. Such a validation is what we attempt here. Classical molecular dynamics simulations of plastic deformation are now possible with even modest computer hardware and can be used to generate in-silico samples of cold worked materials containing complex dislocation networks of varying densities. Analysis of these cells with well-established software tools~\cite{0965-0393-18-1-015012, 0965-0393-18-2-025016} can establish, without ambiguity, the dislocation content of the simulated material. Next, theoretical diffraction patterns and line profiles can be generated using the atomic coordinates and these can be analysed with existing LPA procedures. The inferred details of the dislocation content can then be compared with the ``ground truth'' from the atomistic analysis. This approach, which we describe in detail in the present paper, amounts to the use of simulation to test the validity of the analytical models of lattice strain and its effect on diffraction that are invoked in developing the LPA procedure. This is ``simulation as a test of theory'' \cite{Frenkel:2002aa}; an important role for atomistic simulation that has been somewhat eclipsed by its use as a direct predictive tool.

Though a direct validation of the type described above has not, to our knowledge, been attempted previously, simulations have found use in validating the results of LPA. Zhang et al.\ \cite{Zhang:2020aa} compared estimates of the dislocation content of grain boundaries in nanocrystalline Pd with direct calculations of atomistic models of nanocrystals, finding encouraging agreement. It is important to note that the comparison in Zhang et al.'s work is between different systems: the experimental and the simulation. Kamminga and Delhez \cite{Kamminga:2000aa} used simulations to explore the validity of Wilkens' analytical expressions for the Fourier coefficients of crystals containing a restrictedly-random dislocation distribution. They compared the analytical form with numerically calculated coefficients for the strain field due to a dislocation distribution calculated as a superposition of displacements according to the Nabarro model \cite{Nabarro:1967aa}.

In Section~\ref{sec:methods} we describe in detail the various simulations we have performed and the nature of our analysis, including providing a brief overview of the LPA procedure. In Section~\ref{sec:results} we present and discuss our results. The implications of the present work for LPA are presented in Section~\ref{sec:conclusions}, along with a discussion of probable productive avenues of future study.

\section{Methods}\label{sec:methods}
\begin{figure}\begin{center}
{\includegraphics[width=0.9\figwidth]{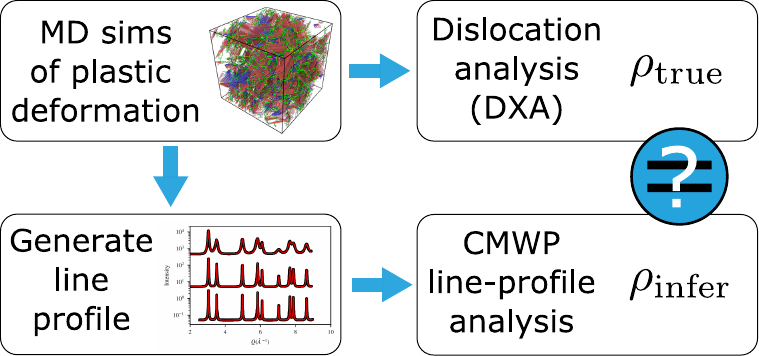}}
\caption{A schematic representation of the process we use to validate the inferences from line profile analysis.}
\label{fig:schematic}
\end{center}
\end{figure}

The process we go through in order to test the inferences in LPA is represented schematically in Figure~\ref{fig:schematic} and is as follows:
\begin{enumerate}
\item we produce atomistic models of material containing representative distributions of dislocations;
\item we analyse these models to determine their precise defect content;
\item we also generate theoretical diffraction patterns from the models;
\item we apply line profile analysis to the diffraction profiles to infer a second measure of defect content.
\end{enumerate}
We will then be able to compare the directly calculated defect content with that inferred from LPA. Each of the above steps is described in more detail below.

\subsection{Creating dislocation distributions}
We have created dislocation distributions in several ways, to ensure that we have a range of behaviour in the shapes of the peaks in the line profiles. In all cases we use a model of copper, represented by an existing embedded atom method (EAM) empirical potential \cite{Ackland:1987hs}. We choose copper because there is abundant experimental data relating to the use of LPA (see for example Refs \cite{Ungar:1984aa,Groma:1988aa,Ungar:1989aa,Balogh:2006aa}). Dynamic simulations and static relaxations are carried out with the Lammps software \cite{Plimpton:1995fv}.

The defect production methods are discussed below, but we note now that, in general, it is not of critical importance that the dislocation distributions be strictly realistic. The approach we are taking to validating LPA compares the true defect statistics with the statistics inferred from the diffraction profiles \emph{for the same cells}. Clearly we do not wish to study systems that are too far removed from reality, but to the extent that our artificially produced defective samples differ from those typically seen in experiment, then this perhaps represents a stronger test of the ability of LPA to measure defect content, since the methodology will tend to have been optimised for, and previously tested only on, experimental samples.

\subsubsection{Plastically-deformed material}
Our first class of test material is ``cold-worked copper''. Our models consist of cubes of face-centred cubic perfect crystal of side length $\SI{361}{\angstrom}$ containing $\sim4$ million atoms, with periodic boundary conditions in all three dimensions, ``seeded'' with six $1/2\langle 110 \rangle$-type dislocation loops of $\SI{12}{\angstrom}$ radius at random locations. Following an initial $10\,\tt{ps}$ anneal at $300\,\tt{K}$ (with a Nose-Hoover thermostat and anisotropic zero-pressure barostat), each of these cells is deformed by first compressing uniaxially to a prescribed percentage reduction in thickness and then compressing biaxially in the other two directions to return the sample to an approximately cubic shape. The compressions are implemented by applying a fixed engineering strain rate in the chosen direction (using the `deform' fix in Lammps) and applying zero-pressure conditions in the other dimension(s) via Nose-Hoover barostats.  A final anisotropic relaxation of the size of the simulation cell and the atomic positions is then carried out. The chosen compression fractions are 5\%, 10\%, 20\%, 40\%, 60\% and 80\%. The two stages of the compression each last $0.5\,\tt{ns}$ and are carried out under the action of a Nose-Hoover thermostat at $300\,\tt{K}$. We have previously used this method to generate deformed material for investigations of dislocation-driven grain boundary motion~\cite{Race:2019ac}.

Clearly the strain rates applied here are colossally high ($2\times 10^8\,\tt{s}^{-1}$ - $1.6\times 10^9\,\tt{s}^{-1}$), and are orders of magnitude beyond experimentally achievable strain rates. However, as pointed out in~\cite{Race:2019ac}, the sample size is very small and so the maximum strain rate implies a velocity of $\sim 60\, \tt{m}\,\tt{s}^{-1}$ for an atom at the surface, which corresponds to an achievable, though high, rate of strain in a macroscopic sample.
The material produced by this process will be denoted as ``cold-worked'' or ``CW''.

We then carry out a process of ``annealing'', by running further dynamics on the cold-worked samples for a duration of $0.5-2.5\,\tt{ns}$ at a temperature of $800\,\tt{K}$. Again, computational constraints mean we can afford only short simulations, but as we show in Ref~\cite{Race:2019ac} this short heat treatment is long enough for the early stage relaxation of the dislocation distributions to take place.
We will refer to this heat treated material as ``annealed''.

\subsubsection{Restrictedly-random dislocation distributions}
In an effort to bring physical meaning to the outer cut-off radius $R_e$ invoked in line profile analysis, Wilkens introduced the idea of a \emph{restrictedly-random distribution} of dislocations~\cite{Wilkens:1969ab,Wilkens:1969ac}. Broadly speaking we consider that the overall sample of material is sub-divided into regions, which we term \emph{cells}, each having the same dislocation density and in each of which the dislocations are randomly distributed in space. As the size of the cells is reduced, the dislocation distribution is forced to become more homogenous and more strongly correlated.

To explore the notion of $R_e$, and the equivalent parameter $M$ introduced by Wilkens~\cite{Wilkens:1969ab,Wilkens:1969ac}, we have created atomistic realisations of restrictedly-random distributions of dislocations with different cell sizes. We construct quasi-two-dimensional fully-periodic supercells containing edge dislocations of burgers vector $\vec{b}=\pm 1/2[110]$ with line direction $[001]$. These cells are constructed using a two-atom body-centred tetragonal unit cell defined by lattice vectors $1/2[110]$, $1/2[1\bar{1}0]$ and $[001]$ in the conventional FCC lattice, such they are  $\sim 300\,\tt{nm}$ square, and $3.61\,\tt{\AA}$ in the third dimension, containing $\sim 3$ million atoms. These supercells are divided into square cells of side lengths varying from $12.8\,\tt{nm}$ (i.e.\ a grid of $24\times 24$ cells) up to the size of the entire supercell (i.e.\ a single cell). Each cell is then populated with edge dislocations such that every cell, in every supercell, contains an equivalent dislocation density, corresponding to 1152 dislocations in the whole supercell. We also enforce equal numbers of dislocations with opposite burgers vectors in each cell, so that the smallest cell contains a single dislocation dipole.

To relax the as-constructed configurations we first relax the atomic coordinates using a conjugate gradient algorithm with a maximum force tolerance of $10^{-6}\,\tt{eV}\,\tt{\AA}^{-1}$, then anneal the supercell at $600\,\tt{K}$ with a Nose-Hoover thermostat for $4\,\tt{ps}$ (2000 steps of $2\,\tt{fs}$), before finally relaxing the atomic configuration and now also the supercell size (to zero external stress) with the same force tolerance as previously. We have optimised the annealing time to be as short as possible to allow good relaxation, but without allowing the dislocation distribution to evolve unduly. Nevertheless, significant movement of dislocations occurs in some cases, as discussed in Section~\ref{sec:results:const-random}.

\subsection{Calculating defect content from atomistic data}
To calculate the defect content of our simulated dislocation distributions we make use of the Dislocation Analysis tool (DXA) \cite{0965-0393-18-2-025016} in the Ovito software~\cite{0965-0393-18-1-015012}. Ovito undertakes a comprehensive burgers circuit analysis of the set of atomic coordinates and returns a list of dislocation segments, giving position, line length and direction, and burgers vector for each. From this output we are then able to determine the densities of dislocations present for dislocations of various types and characters. We have tested the sensitivity of the calculated line density to the parameters of the DXA algorithm and found less than 10\% variation in the results for significant variation of the parameters about their default values.

\subsection{Calculating diffraction patterns}
We have adopted two approaches to calculating the diffraction patterns for our simulated distributions of defects and these will be defined below. In both cases we consider the scattering of an incoming x-ray beam with wavevector $\k$ into an outgoing beam with wavevector $\k'$ by an array of atoms with positions $\{ \rr_i\}$ indexed by $i$. The relative phase difference for the scattered wave from the $i^{\tt{th}}$ atom will be $\exp(-\mi \Q\cdot\rr_i)$, where $\Q=\k'-\k$ and $\hbar\Q$ is the momentum transfer. We then write the amplitude of the scattered wave in the direction of $\k'$ (assuming a fixed $\k$) as
\eq\label{eq:scattered_amplitude}
\Psi(\Q) =\sum_i \me^{-\mi  \Q\cdot \rr_i},
\qe
where we have omitted a prefactor which will affect only the normalisation of the final diffraction patterns and we are assuming that all the atoms in our sample are of the same type and the angular variation in scattered amplitude can be neglected.

\subsubsection{Real-space method}\label{sec:diffraction:realspace}
To calculate the full line profile we consider the scattered \emph{intensity} corresponding to~\eqref{eq:scattered_amplitude}: $I(\Q)=\Psi^{\ast}(\Q)\Psi(\Q) = \sum_{i,j} \me^{-\mi  \Q\cdot \rr_{ij}}$, where $\rr_{ij}=\rr_j-\rr_i$. To obtain the equivalent of a powder diffraction pattern (or a pattern from an untextured polycrystal) we must integrate this expression over the full range of solid angle
\eq
I(Q) = \int_{0}^{2\pi} \md \phi \int_{0}^{\pi}  \md \theta \sin\theta \, \sum_{i,j} \me^{-\mi \Q\cdot \rr_{ij}},
\qe
where we are assuming that for each atom pair the vector $\rr_{ij}$ lies along the zenith direction ($\theta=0$) of the polar coordinates. Completion of the angular integrals leads to the \emph{Debye equation} \cite{Debye:1915aa}
\eq\label{eq:debye}
I(Q) = 4\pi \sum_{i,j}\frac{\sin (Q\,r_{ij})}{Q\,r_{ij}},
\qe
which allows us to calculate the whole line profile $I(Q)$ from a set of atomic coordinates.

Because the dependence on the atomic coordinates is only through the interatomic separations $r_{ij}=|\rr_{ij}|$, a practical approach to implementing a calculation using~\eqref{eq:debye} is to first calculate a histogram of the interatomic separations for all pairs of atoms to arrive at a set of bin centres and frequencies $\{(R_s, f_s) \}$. The diffraction profile is then given by
\eq
I(Q) = 4\pi \sum_{s}f_s\,\frac{\sin (Q\,R_s)}{Q\,R_s} \bigg/ \sum_{s}f_s.
\qe
The best results are obtained by including only atoms within a sphere circumscribed within the full simulation supercell. If the full cuboidal supercell is considered then the diffraction peaks corresponding to reciprocal lattice vectors parallel to the supercell edges show prominent thickness fringes which complicate interpretation of the pattern. Considering a sphere of material largely eliminates these fringes.

Calculation of the histogram of interatomic separations scales as $N^2$ for an $N$-atom system and so can be time consuming. Fortunately the calculation is trivially parallelisable and so can be easily implemented on any available multi-processor system. We made use of idle cycles on a Condor pool of desktop computers in the University of Manchester's teaching laboratories and a set of scripts suitable for this purpose is available \cite{Race:tb}.

\subsubsection{Reciprocal-space method}\label{sec:diffraction:reciprocalspace}
In some cases we are unable to make use of the above real-space method (see Section	~\ref{sec:results:const-random}) and instead calculate details of the scattered amplitude in reciprocal space in the vicinity of a particular lattice reflection. For the reflection from the $(hkl)$ planes corresponding to the reciprocal lattice vector $\G_{hkl}$ Equation~\eqref{eq:scattered_amplitude} can be written
\eq\label{eq:reciprocal_space}
\Psi_{hkl}(\q) = \sum_i\me^{-\mi (\G_{hkl}+\q)\cdot\rr_i},
\qe
where $\q$ is a small vector in reciprocal space defined by $\Q = \G_{hkl}+\q$. The one-dimensional profile in $Q$ can then be calculated by integrating $I_{hkl}(Q) = \int_{\Omega} \md\q\,(\Psi_{hkl})^{\ast}(\q)\Psi_{hkl}(\q)\delta(|\G_{hkl}+\q|-Q)$, where $\Omega$ is a volume of reciprocal space encompassing the diffraction spot, $\delta$ is the Dirac delta function and we are ignoring normalisation of the peak.

To implement this calculation practically we define a set of values $\{ \q_{\alpha} \}$ around $\q=0$ and carry out the sum over atoms in~\eqref{eq:reciprocal_space} for each point in reciprocal space $\G_{hkl}+\q_{\alpha}$. The computational cost of this process scales only linearly in the number of atoms $N$, but also scales linearly with $N_q$, the number of values of $\q_{\alpha}$. Because we are interested in details of the \emph{shape} of the diffraction peaks, rather than just the peak positions and widths, a dense sampling of reciprocal space is required and $N_q$ is thus large. Again, calculation of~\eqref{eq:reciprocal_space} is trivially parallelisable over the values of $\q_{\alpha}$.

\subsection{Calculating defect content via line-profile analysis}\label{sec:methods:lpa}
X-ray diffraction line profile analysis relies on the fact that the diffraction peaks for a sample of crystalline material change shape when the crystallites are smaller and when they contain defects. Essentially, the scattered intensity for a given diffraction peak, indexed by $hkl$, as a function of the scattering wave-vector ${Q}$ can be considered as a convolution of the effects of size and strain broadening:
\eq\label{eqn:whapproach}
I_{hkl} = I_{hkl}^{\tt{size}} \ast I_{hkl}^{\tt{strain}}.
\qe
These two causes of broadening can be separately determined because they each have a different dependence on diffraction order~\cite{Warren:1952aa}. The Williamson-Hall method \cite{Williamson:1953aa} approaches this problem by considering the full width at half maximum and integral breadth of each diffraction peak. Whilst this approach is comparatively straight-forward to apply, it reduces the change in shape of the peaks to two parameters, with the result of significant information loss. An alternative approach, the Warren-Averbach method~\cite{Warren:1950aa,Warren:1959aa}, instead examines the Fourier coefficients of the peaks, retaining full information about peak shape. Writing the Fourier variable as $L$, \eqref{eqn:whapproach} becomes
\eq\label{eqn:waapproach}
A_{hkl}(L) = A_{hkl}^{\tt{size}}(L)\, A_{hkl}^{\tt{strain}}(L),
\qe
where $A$ denotes the Fourier coefficients. This decomposition of the broadening can be further extended to consider multiple types of defect as sources of lattice strain, writing, for example,
\eq\label{eq:straindecomp}
A_{hkl}^{\tt{strain}}(L) = A_{hkl}^{\tt{disl}}(L)\, A_{hkl}^{\tt{SF}}(L),
\qe
where the terms on the right-hand side are contributions from dislocations and stacking faults, but other planar defects, inter-granular strains and instrumental effects can also be included. Successful decomposition of the contributions to the strain from multiple defect types relies on the corresponding Fourier coefficients having different patterns of variation from reflection to reflection ($hkl$-dependence) and with diffraction order.

\subsubsection{Convolutional multiple whole profile (CMWP) approach}\label{sec:methods:lpa:cmwp}
A number of groups have developed implementations of the Warren-Averbach method \cite{Delhez:1976wh,Berkum:1994ts,Balzar:2004tg}, but we will focus here on the convolutional multiple whole profile (CMWP) approach developed by Ungar and co-workers~\cite{Ungar:2001aa,Ungar:1999aa,Ungar:1984aa}, which we apply as part of the present work. In essence, CMWP relies upon physical profile functions for the Fourier coefficients $A_{hkl}(L)$ for different defect types. These profile functions are parameterised in terms of the density and character of the defects in the material. It then remains to optimise the values of these parameters such that they give rise to a theoretical line profile that best matches the experimental data.

The above brief explanation neglects much complexity and subtlety in the approach and we refer the reader in particular to Ungar et al.\ \cite{Ungar:2001aa} and Ribarik et al.\ \cite{Ribarik:2020aa} for more detailed accounts. For our purposes we need only consider some small details of the behaviour of the profile functions associated with dislocations.

Warren and Averbach \cite{Warren:1950aa} showed that the effect of strain gives rise to Fourier coefficients of the form
\eq\label{eq:wa14}
A_{hkl}^{\tt{disl}}(L) \approx \exp (-G_{hkl}^2 L^2\langle \varepsilon^2 \rangle_L /2),
\qe
where $G_{hkl}$ is the centre of the diffraction peak and $\langle \varepsilon^2 \rangle_L$ is the mean-square strain measured between pairs of points in the material separated by a distance $L$ in a direction perpendicular to the $(hkl)$ planes. This form for $A_{hkl}^{\tt{disl}}(L)$ also assumes that we consider only small values of $L$ and small strains. The derivation of \eqref{eq:wa14} is rather involved and can be found in Ref \cite{Warren:1950aa} in which \eqref{eq:wa14} appears as Equation 14.

Taking the logarithm of \eqref{eqn:waapproach} then gives us a form for the Fourier coefficient of the diffraction peaks
\eq\label{eqn:logAfull}
\ln A_{hkl}(L) = \ln A_{hkl}^{\tt{size}}(L)-\frac{1}{2}G_{hkl}^2 L^2\langle \varepsilon^2 \rangle_L.
\qe

In the formal application of the Warren-Averbach approach \cite{Warren:1950aa,Warren:1959aa,Delhez:1976wh,Berkum:1994ts,Balzar:2004tg} the mean-square strain is assumed to be a constant independent of $L$. All experiments show, however, that the mean-square strain is an $L$-dependent quantity in accordance with its being a correlation function showing the dislocation's strain correlation in crystal space. In order to find the physically correct strain correlation a specific lattice defect needs to be considered. In the original work of Warren and Averbach \cite{Warren:1950aa} and Warren \cite{Warren:1959aa} defect-independent `general' strain was invoked. With such an assumption it is difficult or almost impossible to derive the correct $L$ dependence of the mean-square strain. In \cite{Warren:1950aa} and  \cite{Warren:1959aa} two extreme strain correlation functions were considered. In one a random uncorrelated strain distribution is assumed and in the other a hyperbolic L dependence. Neither of these two assumptions are substantiated for any real lattice defect. Krivoglaz \cite{Krivoglaz:1996ug, Krivoglaz:1963aa} considered dislocations to derive the $L$ dependence of $\langle\varepsilon ^2\rangle_{L}$. He showed that for small $L$ values the strain correlation function decays logarithmically with $L$. The logarithmic function was given by two parameters: the dislocation density, $\rho$, and the crystal size, $D$. Later Groma \cite{Groma:1988aa} showed that the logarithmic $L$ dependence of the strain correlation function is the direct consequence of the invariance in the strain correlation of dislocations against linear stretching, see e.g. Equation 35 in Ref \cite{Groma:1988aa}. Wilkens \cite{Wilkens:1969ab,Wilkens:1969ac} recognised that the crystal size, $D$, in the strain correlation function causes a similar logarithmic singularity in line broadening to that in the elastically stored energy of dislocations. 

In order to eliminate this logarithmic singularity, Wilkens \cite{Wilkens:1970aa} gives a form for the mean square strain for a restrictedly-random distribution of dislocations
\eq\label{eqn:msqstrain}
\langle \varepsilon^2 \rangle_L \approx
\frac{\rho C b^2}{4\pi}\ln\left( \frac{R_{\tt{e}}}{L} \right),
\qe
where $\rho$ is the dislocation density, $b$ is the length of the burgers vector and $R_{\tt{e}}$ is the size of the ``cell'' for the restrictedly-random distribution. $C$ is an average contrast factor, which quantifies the strength of the effect of a given dislocation type on the diffraction peak under study. The fundamental origins of the Krivoglaz-Wilkens strain function in ~\eqref{eqn:msqstrain} are detailed in Refs.~\cite{Groma:1997tk,Groma:2016uc,Zaiser:2001ur}.

The $L$ dependence of the SCF in \eqref{eqn:msqstrain} is still only valid for small values of $L<R_{\tt{e}}$. In practical terms this means that \eqref{eqn:msqstrain} only gives the tail regions of diffraction peaks correctly. Wilkens also calculated the SCF in the entire L range using the restrictedly-random dislocation distribution concept and derived a general strain function $f(\eta)$ with $\eta=L/R_{\tt{e}}$ \cite{Wilkens:1969ac} so that \eqref{eqn:msqstrain} becomes 

\eq\label{eqn:msqstrainmod}
\langle \varepsilon^2 \rangle_L \approx
\frac{\rho C b^2}{4\pi}f(\eta).
\qe

The first part of $f(\eta)$, for $\eta<1$, is logarithmic, as in \eqref{eqn:msqstrain}, whereas the second part, for $\eta>1$, is hyperbolic. The full $f(\eta)$ function is given in eqs. (A.6)-(A.8) in \cite{Wilkens:1969ac}. The form given in \eqref{eqn:msqstrainmod} is used in the CMWP method to infer defect density. As mentioned in relation to \eqref{eq:straindecomp}, it is the fact that the contrast factors $C$ are a function of $hkl$ and that $C$ and $R_{\textrm{e}}$  also show different behaviour for different defects (including for dislocations on different slip systems) that allows the CMWP method to disentangle the contributions to lattice strain from multiple defect types. We make use of this in the analysis of our results.

In Section~\ref{sec:results:const-random} we will use \eqref{eqn:logAfull} to understand the meaning of $R_{\textrm{e}}$ in the context of a restrictedly random distribution of dislocations. In doing so we will focus on a single Bragg reflection corresponding to a set of planes parallel with the edges of a crystal of square geometry. Warren and Averbach point out that the initial slope of the Fourier coefficients (at $L=0$) will be equal to the reciprocal of the crystal size parallel to $\vec{G}_{hkl}$ (the ``mean column length'' in their analysis). In the case of the crystal geometry that we will analyse, in which all the ``columns'' have the same length, the relationship is even stronger, enabling us to write
\eq \label{eqn:sizeeffect}
A_{hkl}^{\tt{size}}(L) = 1 - \alpha L,
\qe
in which $\alpha = 1/D$, for $D$ the size of the crystal in our simple geometry. The form of this relationship emerges from elementary diffraction theory. For our sample geometry and choice of diffraction peak in Section~\ref{sec:results:const-random} we can consider a single slit with a width corresponding to the crystal size. The Fourier transform of this ``top hat'' function is an amplitude that is a $\mathrm{sinc}$ function giving an intensity that varies as $\mathrm{sinc}^2$. Transforming this intensity back into real-space gives a triangular peak, falling off linearly about $L=0$.

This now allows us to write the general form of \eqref{eqn:logAfull} as
\begin{align}\label{eqn:logAsimple}
\ln A_{hkl}(L) &= \ln(1-\alpha L) - \beta \frac{\rho}{\rho_0} L^2 \ln\left( \frac{R_{\tt{e}}}{L} \right), \nonumber
\\
\alpha &= 1/D, \nonumber
\\
\beta &= \frac{G^2_{hkl}\rho_0 C b^2}{8\pi}
\end{align}
in which we have introduced a reference dislocation density $\rho_0$. We will make use of \eqref{eqn:logAsimple} in Section~\ref{sec:results:const-random} treating $\alpha$, $\beta$ and $R_{\tt{e}}$ as fitting constants. We note that when $\alpha L$ is small we can further approximate the first term of~\eqref{eqn:logAsimple} as  $\ln(1-\alpha L)\approx -\alpha L$.

The length-scale $R_{\tt{e}}$ in \eqref{eqn:msqstrain} has a clear physical interpretation in the context of Wilkens' restrictedly-random distributions of dislocations. An alternative, dimensionless, parameter, also due to Wilkens, is often also quoted and is defined,
\eq
M = R_{\tt{e}}\sqrt{\rho}.
\qe
This can be interpreted as measuring the ``dipole character'' of the dislocation distribution, with small values of $M$ corresponding to highly correlated distributions of dislocations with strong dipole character. This interpretation of $M$ is consistent with Wilkens' definition of $R_{\tt{e}}$: as pairs of dislocations with opposing burgers vectors are brought into close proximity the range of their combined strain field is reduced. Conversely, an entirely uncorrelated distribution of dislocations (large $M$) corresponds to the case of large $R_{\tt{e}}$.

\section{Results and discussion}\label{sec:results}
We will now consider our two types of artificial defective material in turn, analysing the defect content, exploring the diffraction patterns and comparing inferences from these patterns with the true defect densities.

\subsection{Cold-worked and annealed material}
\subsubsection{Defect content of simulation cells}
\begin{figure*}\begin{center}
\subfloat[10\%]{\includegraphics[width=0.32\widefigwidth]{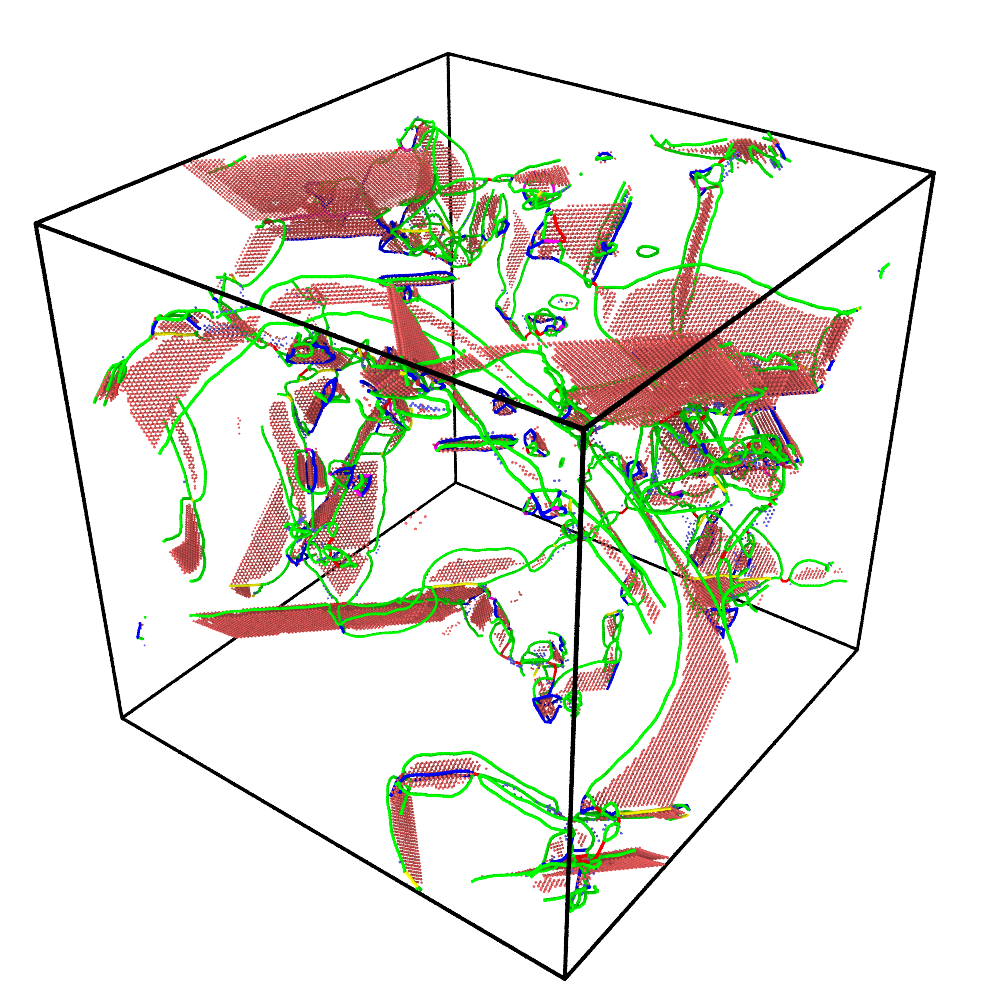}}
\subfloat[40\%]{\includegraphics[width=0.32\widefigwidth]{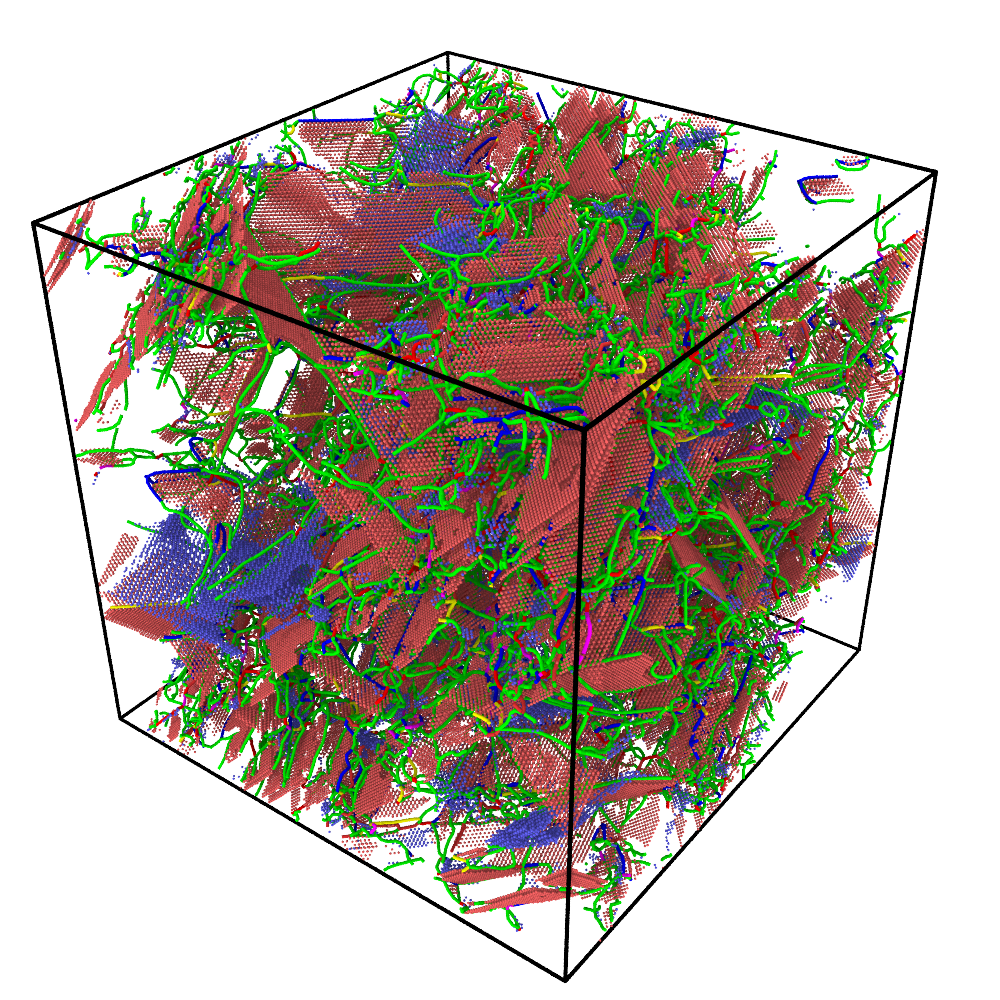}}
\subfloat[80\%]{\includegraphics[width=0.32\widefigwidth]{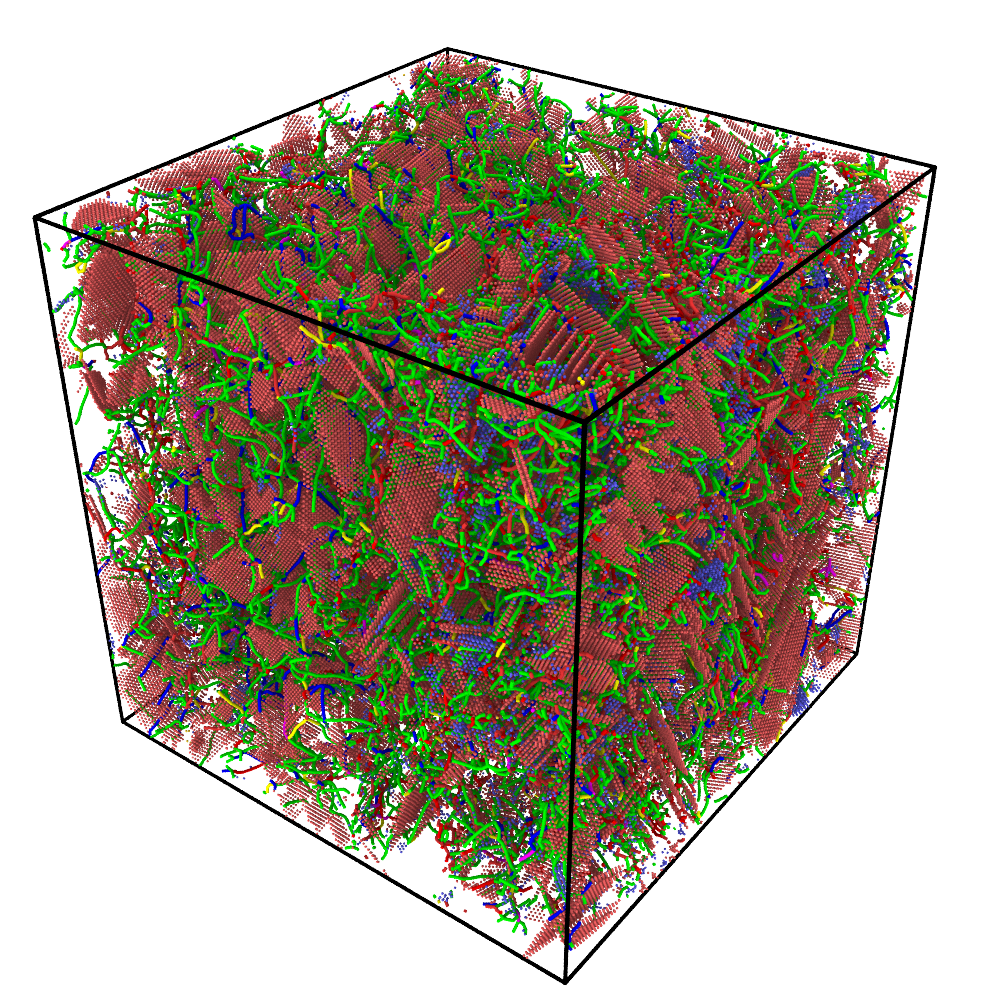}}
\caption{Visualisations of the relaxed defect structures in three of the cold-worked cells at the strains indicated. Dislocations are shown as lines, with colour indicating burgers vector as: $1/6\langle112\rangle$ (green), $1/6\langle110\rangle$ (blue), $1/3\langle001\rangle$ (yellow), $1/2\langle110\rangle$ (red), $1/3\langle111\rangle$ (purple). Atoms in normal (FCC) coordination are not shown. Atoms in HCP coordination are shown in red and those in BCC coordination in blue. Defect analysis was carried out in Ovito~\cite{0965-0393-18-1-015012}.}
\label{fig:coldworked-ovito}
\end{center}
\end{figure*}

\begin{figure}\begin{center}
{\includegraphics[width=\figwidth]{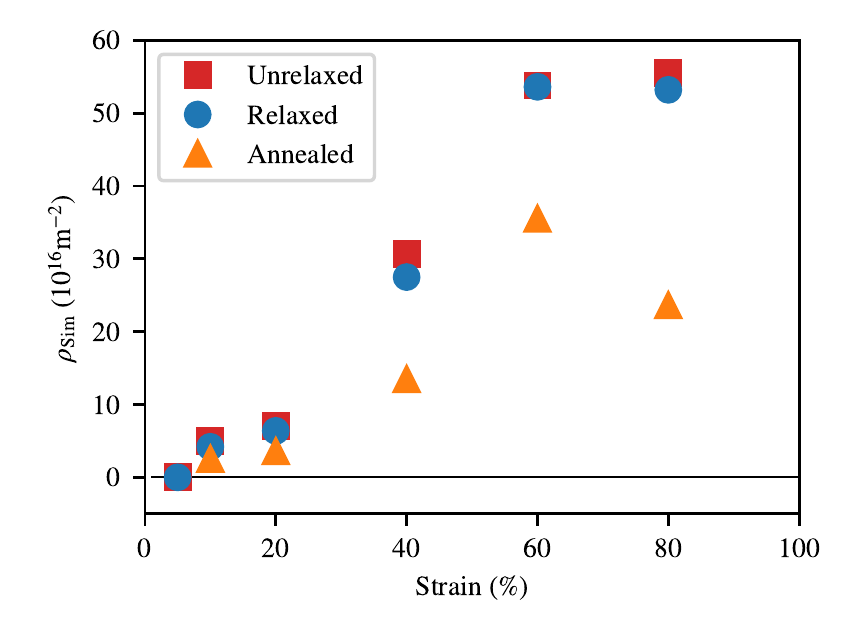}}
\caption{Dislocation content of the cold-worked cells as a function of strain for the as-worked (unrelaxed), relaxed and annealed ($0.5\,\tt{ns}$) cases.}
\label{fig:coldworked-simdensity}
\end{center}
\end{figure}

Figure~\ref{fig:coldworked-ovito} shows visualisations of the defect content for three of the strains explored in the cold-worked material after an initial relaxation. The dislocation densities as calculated using the DXA algorithm in Ovito~\cite{0965-0393-18-1-015012} are shown in Figure~\ref{fig:coldworked-simdensity} and in Table~\ref{table:dislocationdensities} for each cold-work strain in the unrelaxed state (immediately following the biaxial strain), a relaxed state (following an initial relaxation of the atomic coordinates and the supercell shape) and after a $0.5\,\tt{ns}$ anneal at $800\,\tt{K}$. 5\% strain produced no significant multiplication of dislocations from the initially seeded dislocation loops. By 10\% strain the cell already contains a significant density of $ \vec{b}=1/6\langle 112\rangle$ dislocations and several large-area stacking faults. The dislocation density is already significantly beyond that typical in real cold-worked materials, which we attribute to the high strain rate in our simulations, and continues to rise up to 80\% strain. As shown in Table~\ref{table:dislocationdensities}, the dislocation structure is dominated by $1/6\langle 112\rangle$ Shockley partial dislocations at all strains. We also find that a short anneal of $0.5\,\tt{ns}$ is sufficient for the 
initial relaxation of the dislocation structure to take place, with comparatively small changes taking place from that point up to $2.5\,\tt{ns}$. We note that between 60\% and 80\% strains there is very little change in the dislocation density in the as-worked material and that a $0.5\,\tt{ns}$ anneal actually results in a lower density in the 80\% case than for the 60\%. We speculate that by 60\% strain we have reached a sufficiently high dislocation density that defects self-annihilate as fast as they are formed on further strain, such that we have reached a steady state similar to that reported in high-strain-rate simulations by Zepeda-Ruiz et al.~\cite{Zepeda-Ruiz:2020aa}. Evolution of the detailed defect distribution at constant density within this steady state then perhaps results in a defect distribution more amenable to rapid recovery in the subsequent anneal in the 80\% case.

\begin{table*}
\caption{Dislocation content of the cold-worked cells as a function of strain, broken down by dislocation character. An annealing time of $0.0\,\tt{ns}$ indicates the relaxed cold-worked state, prior to annealing. Data for unrelaxed cells are not shown as they differ little from the relaxed case (see Fig.~\ref{fig:coldworked-simdensity}).}
\centering
\begin{tabular}{c c c c c c c}
\hline\hline
Annealing
& Strain   &\multicolumn{5}{c}{Dislocation densities ($10^{16}\,\tt{m}^{-2}$)} \\
time (ns)
& (\%)  
& $1/6\langle112\rangle$
& $1/6\langle110\rangle$
& $1/3\langle001\rangle$
& $1/2\langle110\rangle$
& $1/3\langle111\rangle$
 \\ [0.5ex] 
&
& (Shockley partial)
& (Stair-rod)
& (Hirth partial)
& (Perfect)
& (Frank partial)
 \\ [0.5ex] 
\hline
0.0 & 5 & 0.0 & 0.0 & 0.0 & 0.0 & 0.0 \\
0.0 & 10 & 3.3 & 0.7 & 0.1 & 0.1 & 0.1 \\
0.0 & 20 & 5.0 & 0.7 & 0.2 & 0.3 & 0.2 \\
0.0 & 40 & 22.2 & 2.0 & 1.3 & 1.5 & 0.6 \\
0.0 & 60 & 43.2 & 2.8 & 3.0 & 3.7 & 1.0 \\
0.0 & 80 & 41.0 & 4.6 & 3.1 & 3.8 & 0.8 \\
\hline
0.5 & 10 & 1.4 & 1.0 & 0.0 & 0.0 & 0.1 \\
0.5 & 20 & 2.6 & 0.8 & 0.1 & 0.1 & 0.0 \\
0.5 & 40 & 11.4 & 0.9 & 0.5 & 0.5 & 0.2 \\
0.5 & 60 & 30.3 & 1.2 & 1.5 & 2.1 & 0.4 \\
0.5 & 80 & 19.9 & 1.0 & 0.7 & 1.8 & 0.3 \\
\hline
0.5 & 20 & 2.6 & 0.8 & 0.1 & 0.1 & 0.0 \\
1.0 & 20 & 2.5 & 0.9 & 0.1 & 0.1 & 0.0 \\
1.5 & 20 & 2.3 & 0.8 & 0.1 & 0.1 & 0.0 \\
2.0 & 20 & 2.4 & 0.8 & 0.1 & 0.1 & 0.0 \\
2.5 & 20 & 2.2 & 0.8 & 0.1 & 0.1 & 0.0 \\
\hline
\end{tabular}
\label{table:dislocationdensities}
\end{table*}

\begin{figure}\begin{center}
{\includegraphics[width=\figwidth]{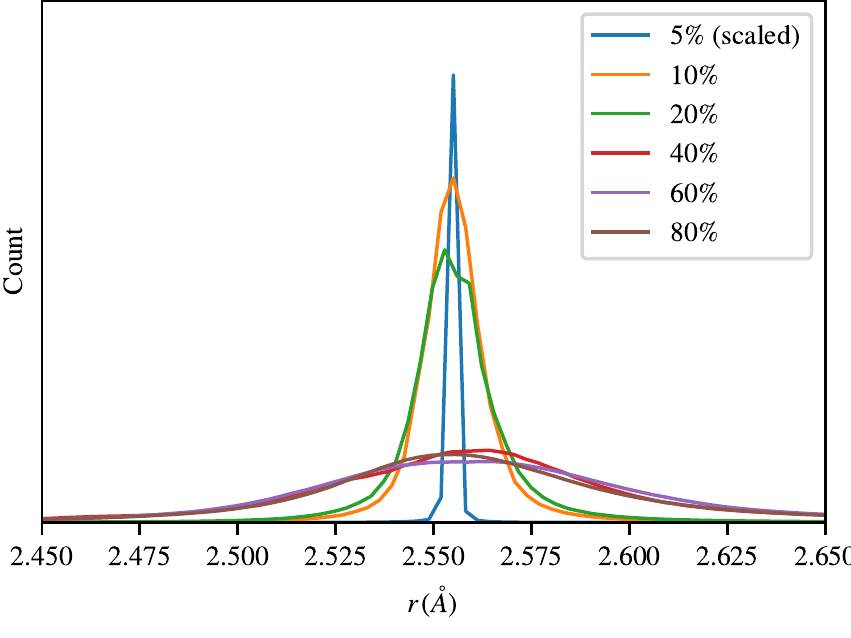}}
\caption{The distribution of pairwise atoms separations around the nearest-neighbour distance in cold-worked and relaxed simulation cells as a function of strain. Note that for the 5\% case, the height of the peak is scaled down by a factor of 4.}
\label{fig:pairwiseseparation}
\end{center}
\end{figure}

Figure~\ref{fig:pairwiseseparation} shows a portion of the pairwise separation histogram, around the nearest-neighbour separation, used to calculate the x-ray diffraction line profile in real space. There is rapid broadening of the distribution from the perfect crystal form up to 40\% strain with comparatively little change thereafter, despite the doubling of dislocation density between 40\% and 60\% strain.

\subsubsection{Calculation of line profiles}

\begin{figure}\begin{center}
\subfloat[]{\includegraphics[width=\figwidth]{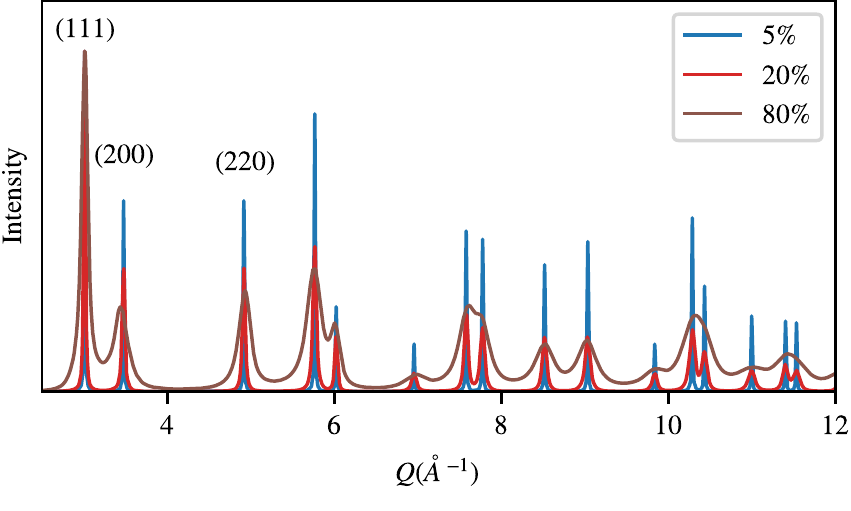}}\hfill
\subfloat[]{\includegraphics[width=\figwidth]{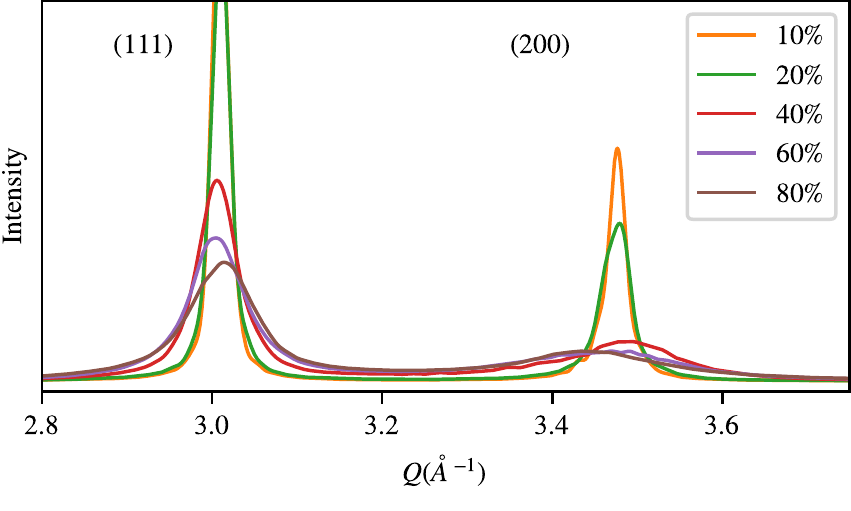}}\hfill
\subfloat[]{\includegraphics[width=\figwidth]{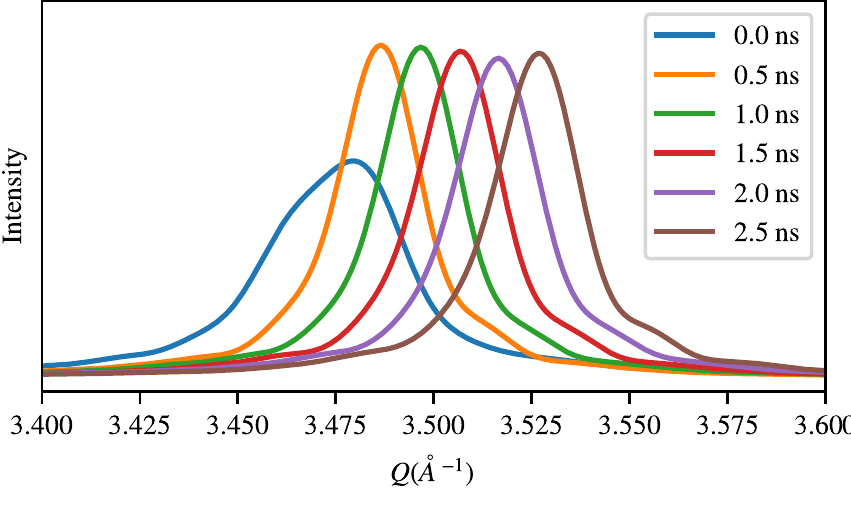}}
\caption{Examples of line profiles for the relaxed cold-worked material calculated using the real-space method. (a) Profiles over a wide range of $Q$ for three strains between the minimum and maximum explored. (b) Detail of two low-index reflections in the profile. (c) The changing shape of the $(200)$ peak as a function of annealing time. A horizontal offset has been applied to aid readability.}
\label{fig:lineprofiles}
\end{center}
\end{figure}

We now discuss the diffraction profiles for the cold-worked material. These are obtained using the real-space approach discussed in Section~\ref{sec:diffraction:realspace}. Figure~\ref{fig:lineprofiles}(a) shows the profiles for three of the applied strains across a moderate range of scattering vector $Q$. The case of 5\% strain is essentially defect free, and so the peaks in this case are broadened only by the finite size of the supercell (size broadening). Note that the broadening in the 80\% case is rather extreme, with significant overlap between adjacent peaks.

Figure~\ref{fig:lineprofiles}(b) shows two of the most prominent peaks for each of the applied strains. Here we can see that significant differences in $(111)$ peak shape are visible between the 60\% and 80\% cases even though the dislocation density in these two cells is very similar.

Figure~\ref{fig:lineprofiles}(c) confirms the data in Table~\ref{table:dislocationdensities}, which showed that the initial effect of annealing is realised within $0.5\,\tt{ns}$, with very little further change in defect density in the next $2\,\tt{ns}$. Here we see the same picture in the shape of the $(200)$ peak which changes significantly early in annealing and then hardly at all.

\subsubsection{CMWP analysis of defect content}
We now have full line profiles for our cold-worked material, to which the process of line profile analysis, as implemented in the CMWP software package, can be applied. This is a direct equivalent of the process that would be used in analysing a profile from an experimental sample. With CMWP we obtain predictions for the dislocation line density and stacking fault fraction in the material. In Section~\ref{results:comparison} we will compare these predictions with our ``ground truth'' derived from atom-by-atom analysis in Ovito \cite{0965-0393-18-1-015012}, but first of all we consider the results of applying the CMWP process in detail.

\begin{figure}\begin{center}
\subfloat[Relaxed structures, several strains.]{\includegraphics[width=0.49\widefigwidth]{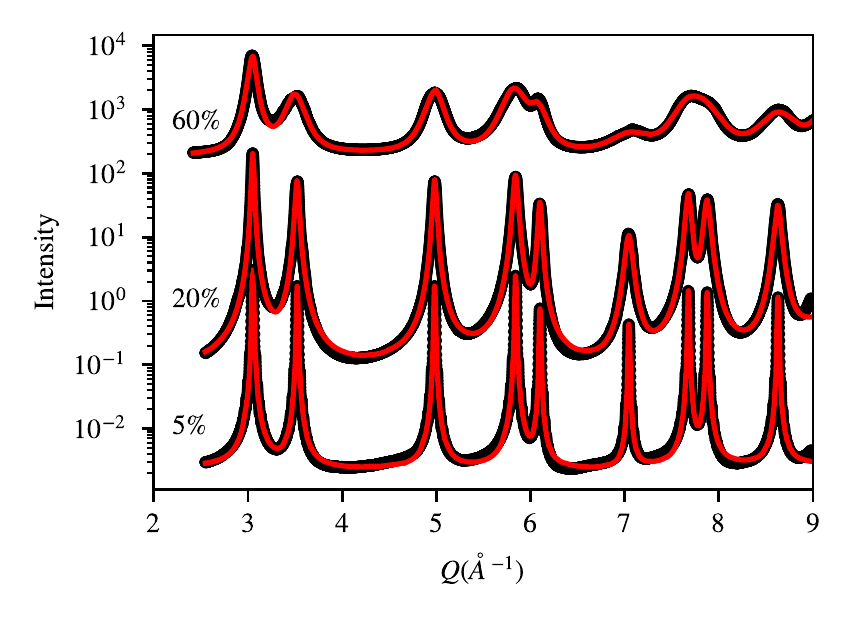}}\hfill
\subfloat[Relaxed and annealed for $0.5\,\tt{ns}$, several strains.]{\includegraphics[width=0.49\widefigwidth]{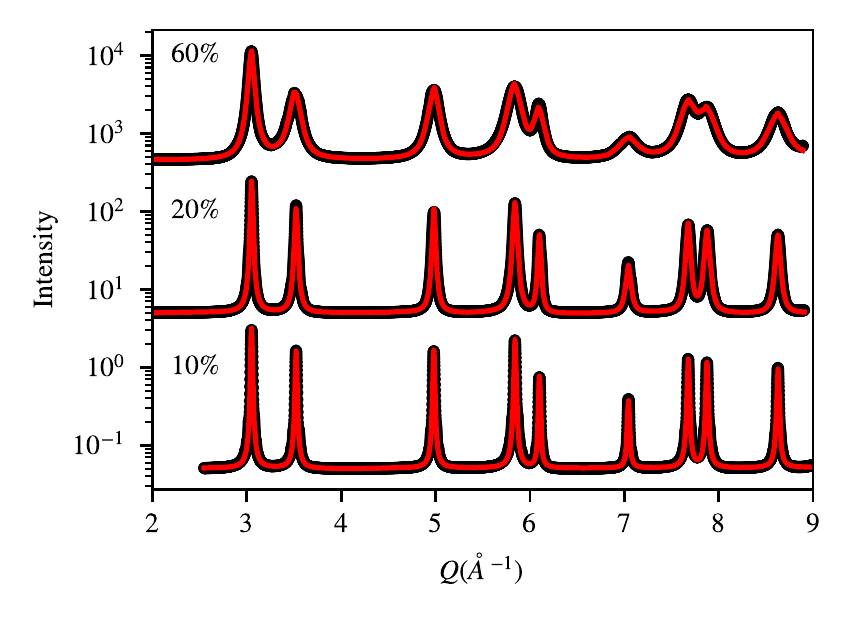}}\hfill
\caption{Comparison of original line profiles (thick, black lines) with CMWP fits (thin, red lines). }
\label{fig:cmwpfits}
\end{center}
\end{figure}

Figure~\ref{fig:cmwpfits} shows a comparison between the original diffraction profiles generated from the simulations cells and the theoretical profiles corresponding to the optimised values of the parameters in the  model employed in the CMWP process. In all cases, for a variety of strain and annealing conditions, these fits are very good, suggesting that there are no features peculiar to the profiles from the simulated material that will prevent successful analysis via the CMWP process.

\begin{figure*}\begin{center}
\subfloat[Dislocation density]{\includegraphics[width=0.45\widefigwidth]{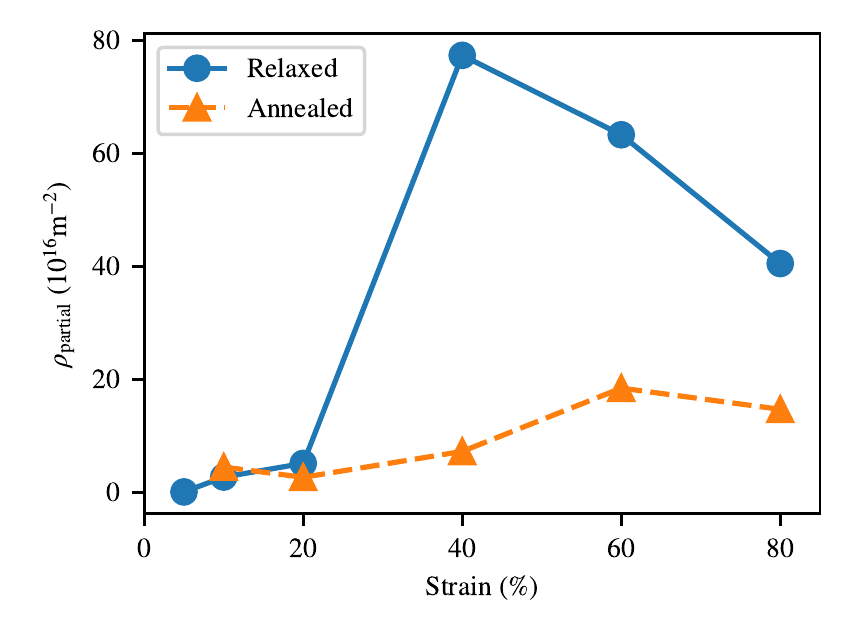}}
\subfloat[Dislocation density by type]{\includegraphics[width=0.45\widefigwidth]{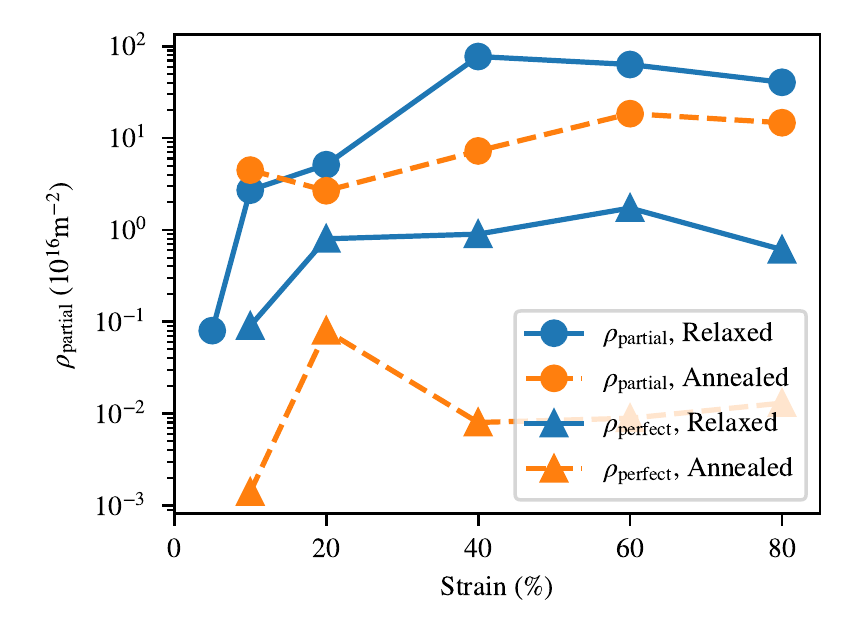}}
\hfill
\subfloat[Stacking fault fraction]{\includegraphics[width=0.45\widefigwidth]{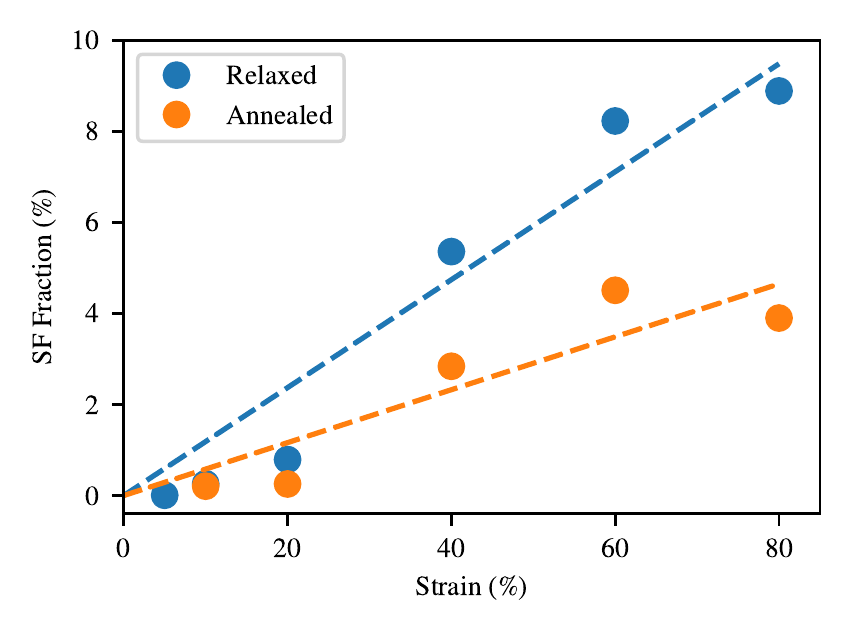}}
\subfloat[Stacking fault spacing]{\includegraphics[width=0.45\widefigwidth]{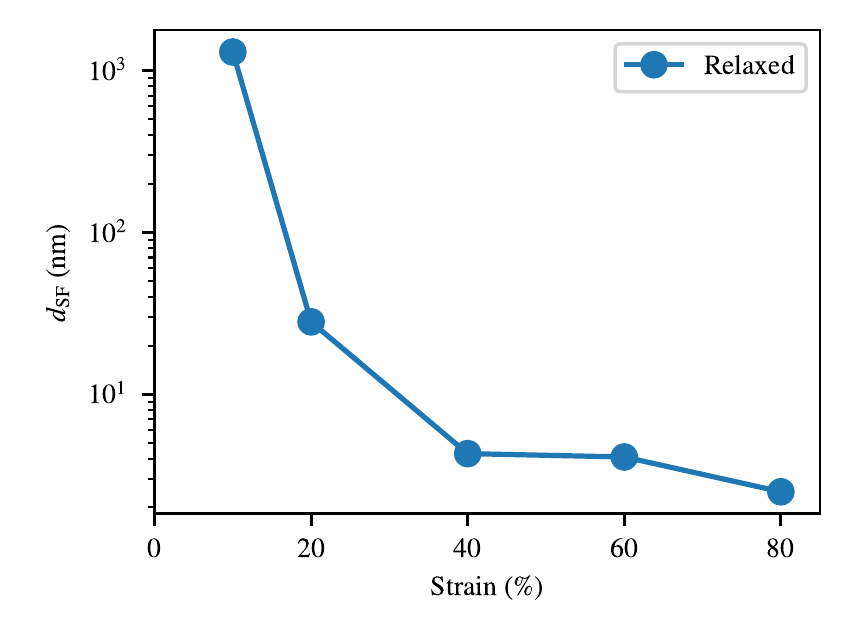}}
\hfill
\subfloat[Outer cut-off radius]{\includegraphics[width=0.45\widefigwidth]{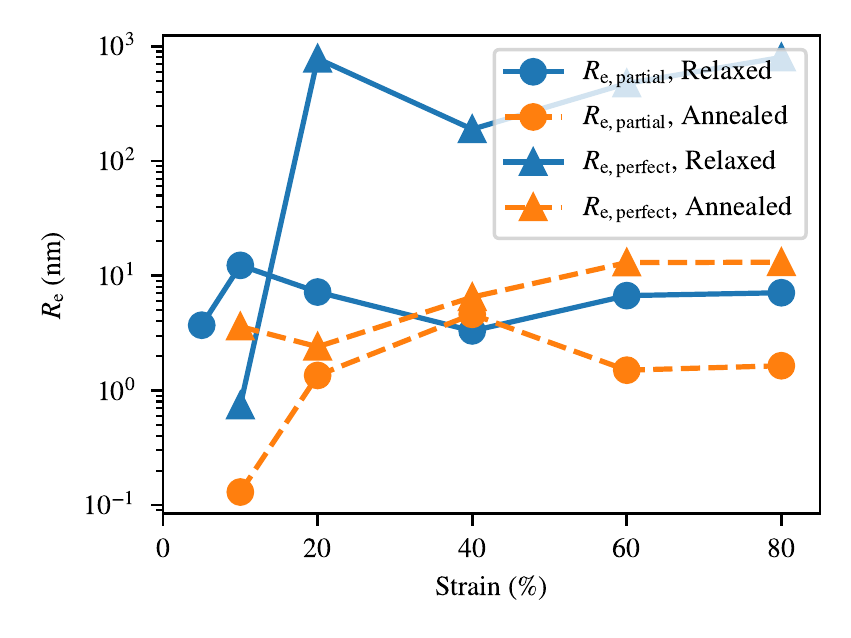}}
\subfloat[Subgrain size]{\includegraphics[width=0.45\widefigwidth]{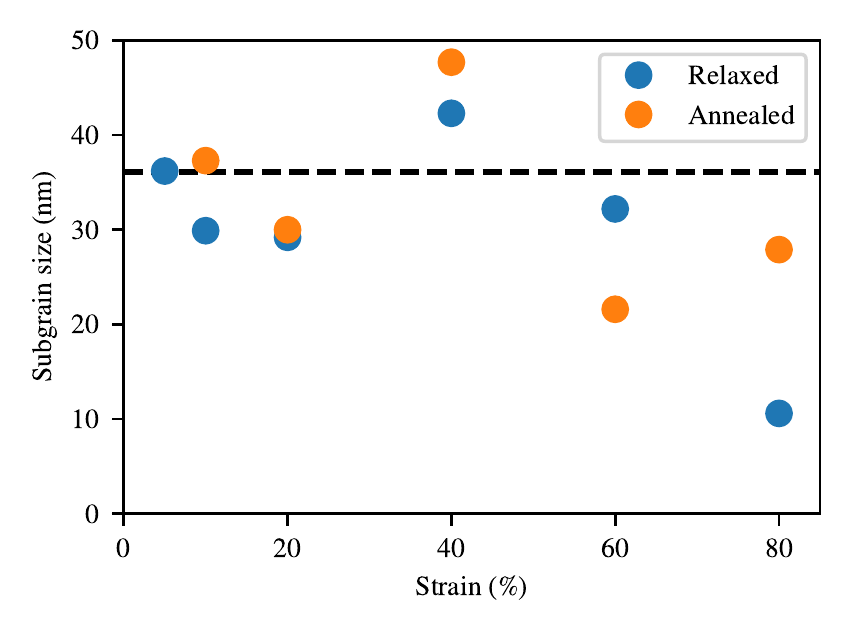}}
\caption{Key parameters for the CMWP analysis, fitted to the x-ray diffraction profiles for the simulation cells. Data for the ``annealed'' case refers to a $0.5\,\tt{ns}$ annealling time. Lines are a guide to the eye, except for the horizontal dashed line in (f), which marks the true size of the grain in the simulated material, for reference.}
\label{fig:cmwpresults}
\end{center}
\end{figure*}

Figure~\ref{fig:cmwpresults} shows the values of several key parameters in the CMWP physical model, optimised in the fitting of the diffraction profiles. 

The CMWP analysis accounts for two dislocation types: the dominant Shockley partials with $\vec{b}=(1/6)\langle 112\rangle$, $b=1.472\,\tt{\\A}$ and perfect dislocations with $\vec{b}=(1/2)\langle 110\rangle$, $b=2.55\,\tt{\\A}$. The dislocation density separated out into these two types is shown in Figure~\ref{fig:cmwpresults}(b).

The stacking fault fraction (Figure~\ref{fig:cmwpresults}(c)) is a measure of the fraction of planes intersected by a line through the crystal that contain a stacking fault at the point of intersection. The stacking fault spacing, $d_{\tt{SF}}$ (Figure~\ref{fig:cmwpresults}(d)), is then given by the reciprocal of this fraction multiplied by the interplanar spacing of the fault plane (in this case the $\{111\}$ planes with spacing $a/\sqrt 3$). 

Figure~\ref{fig:cmwpresults}(e) gives the values of the inferred outer cut-off radius $R_{\textrm{e}}$ in \eqref{eqn:msqstrainmod}. We recall that this parameter relates to the arrangement of the dislocations in the material.

Figure~\ref{fig:cmwpresults}(f) shows the inferred subgrain size in the material. The true size is marked with a line and the good agreement (perfect at 5\% strain) demonstrates how the CMWP approach can distinguish the role of size and strain in broadening the diffraction peaks.

\begin{figure}\begin{center}
\subfloat[]{\includegraphics[width=0.49\widefigwidth]{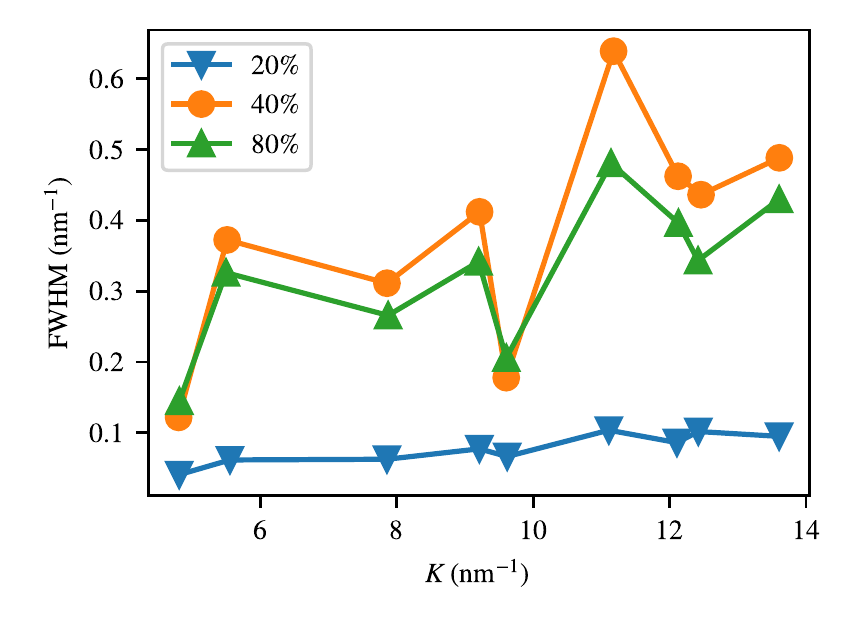}}\hfill
\subfloat[]{\includegraphics[width=0.49\widefigwidth]{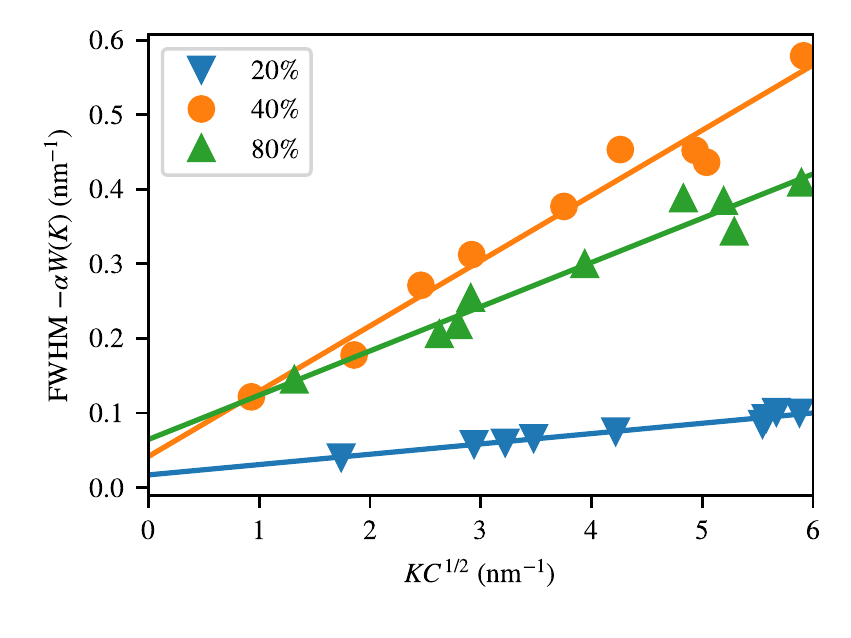}}
\caption{Williamson-Hall (a) and modified Williamson-Hall (b) plots for the simulation cells annealed for $0.5\,\tt{ns}$ as a function of strain.}
\label{fig:whstrain}
\end{center}
\end{figure}

\begin{figure}\begin{center}
\subfloat[]{\includegraphics[width=0.49\widefigwidth]{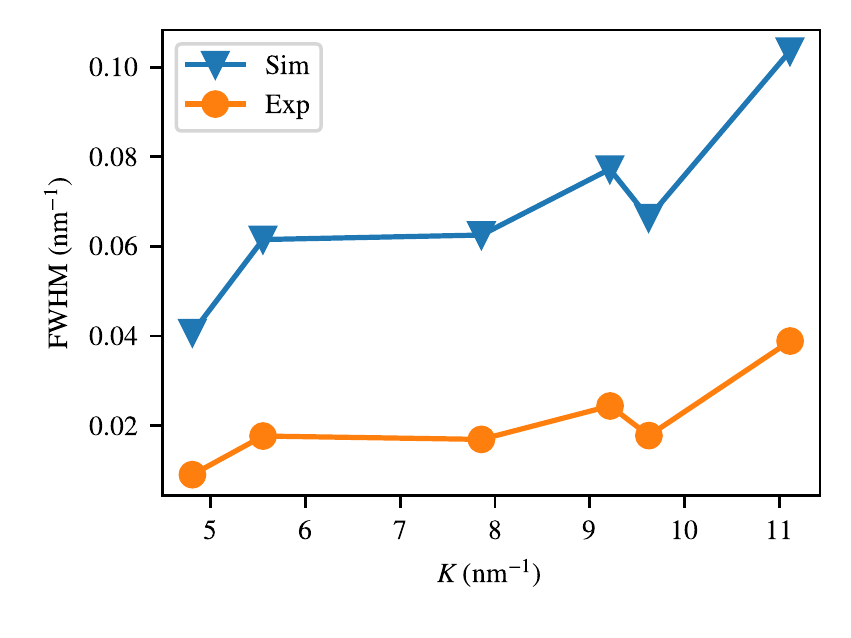}}\hfill
\subfloat[]{\includegraphics[width=0.49\widefigwidth]{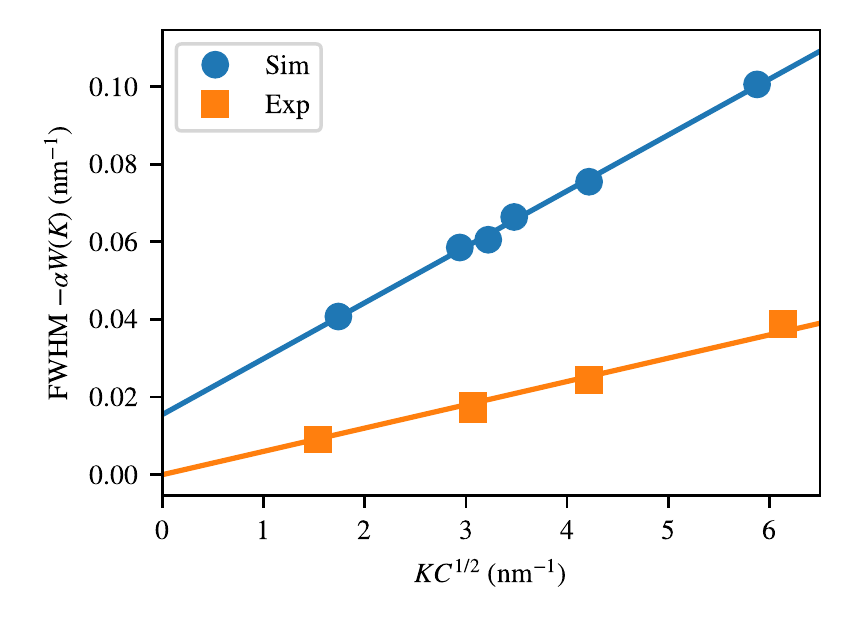}}
\caption{Williamson-Hall (a) and modified Williamson-Hall (b) plots for a simulation cells (20\% strain annealed for $0.5\,\tt{ns}$) compared to experimental data for cold-worked copper. Experimental data are from Ungar and Borbely \cite{Ungar:1996aa}.}
\label{fig:whexp}
\end{center}
\end{figure}

Figures~\ref{fig:whstrain} shows Williamson-Hall and modified Williamson-Hall plots for a range of strains. Figure~\ref{fig:whexp} compares similar plots for one of our simulated patterns to experimental data for ultrafine grained copper crystal, produced by equal-channel angular pressing \cite{Ungar:1996aa} and a similarity in the variation in FWHM across the various peaks is clear. In Figure~\ref{fig:whexp}(b) the non-zero intercept of the modified Williamson-Hall plot for the pattern from the simulation suggests that a substantial fraction of the line broadening is due to size effects (in this case our simulation cell is only $36\,\tt{nm}$ across). This is in contrast to the experimental pattern, for which the intercept is zero and which comes from a sample with a grain size of $\sim 200\,\textrm{nm}$.

\subsubsection{Comparison of CMWP analysis and ``ground truth''}\label{results:comparison}
\begin{figure}\begin{center}
\subfloat[]{\includegraphics[width=\figwidth]{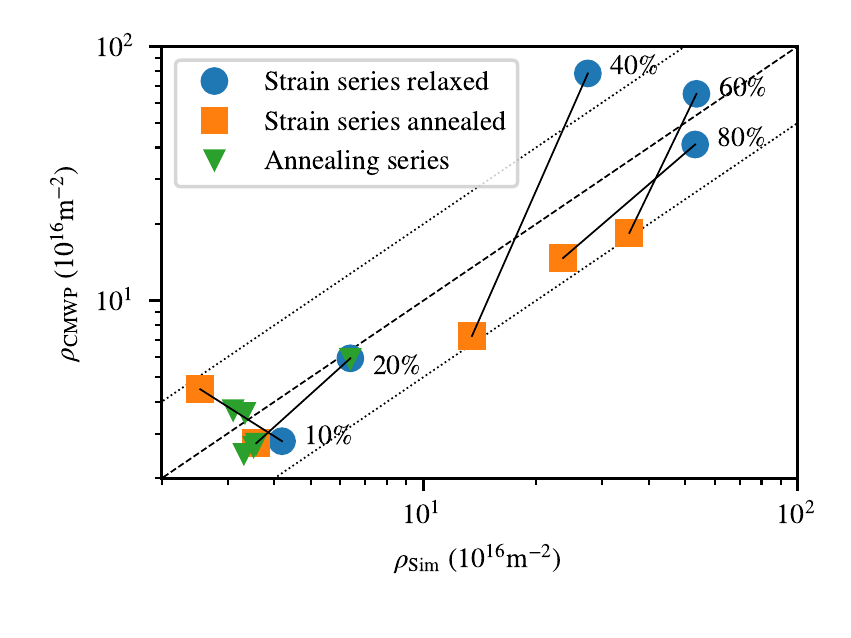}}\hfill
\subfloat[]{\includegraphics[width=\figwidth]{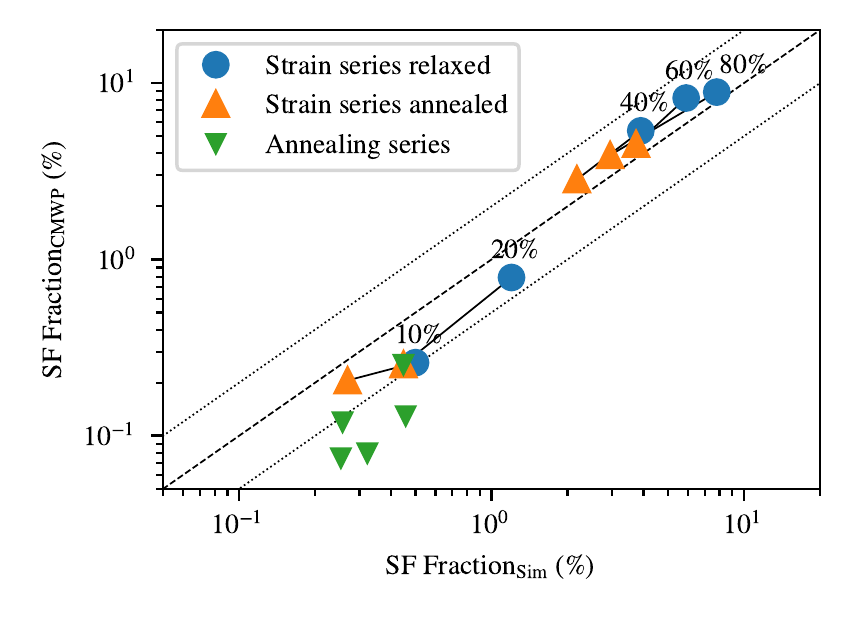}}
\caption{A comparison of (a) dislocation densities and (b) stacking fault fractions derived from line profile analysis with CMWP and the result of atom-by-atom analysis for the relaxed (blue circles) and $0.5\,\tt{ns}$ annealed (orange upright triangles) cells. Solid black lines join relaxed and annealed points corresponding to the same initial strain to aid reading. Data for the 20\% strain case after a variety of annealing times are also shown (green upside-down triangles). The dashed black lines $\rho_{\tt{Sim}}=\rho_{\tt{CMWP}}$ indicates perfect agreement between the atomistic analysis and CMWP, with dotted lines indicating agreement within a factor of 2.}
\label{fig:comparisonrhosf}
\end{center}
\end{figure}

We now come to the key point of this paper, a comparison between the defect content inferred via the CMWP method and the true values. Figure~\ref{fig:comparisonrhosf}(a) shows a comparison between the CMWP dislocation densities and those from simulation cell analysis for the relaxed and annealed cold-worked material. Though quantitative agreement is not perfect, the true dislocation density varies over two orders of magnitude and CMWP does a good job of detecting this variation, generally being within a factor of 2 of the true value.

Figure~\ref{fig:comparisonrhosf}(b) shows a comparison between the CMWP stacking fault fraction 
and the fraction of atoms classified as having local HCP coordination in a common neighbour analysis. We see excellent agreement across almost two orders of magnitude of stacking fault fraction, with large discrepancies only at fractions below a few tenths of a percent. In particular, agreement for the cells annealed for $0.5\tt{ns}$ is excellent across the full range of strains.

These results offer strong support for the use of the CMWP approach for quantifying defects. It is particularly impressive that the method can simultaneously give a quantitative account of stacking faults and dislocations.

\subsection{Restrictedly-random dislocation distributions}\label{sec:results:const-random}
\begin{figure*}\begin{center}
\subfloat[50 unit cells $=12.6\,\tt{nm}$]{\includegraphics[width=0.32\widefigwidth]{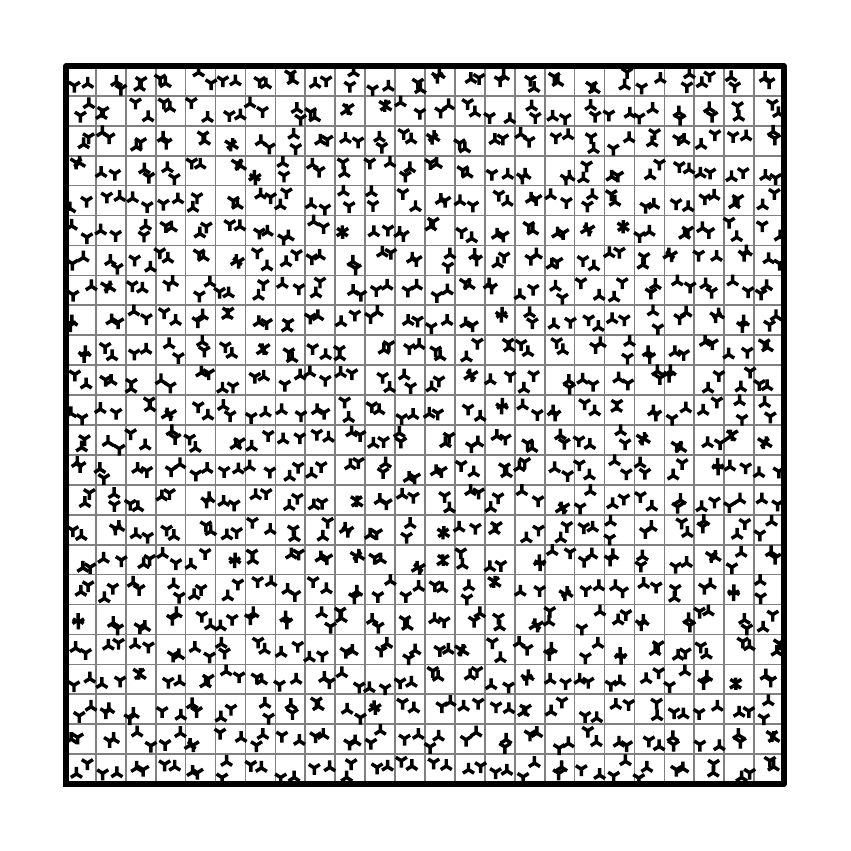}}
\subfloat[200 unit cells $=51.1\,\tt{nm}$]{\includegraphics[width=0.32\widefigwidth]{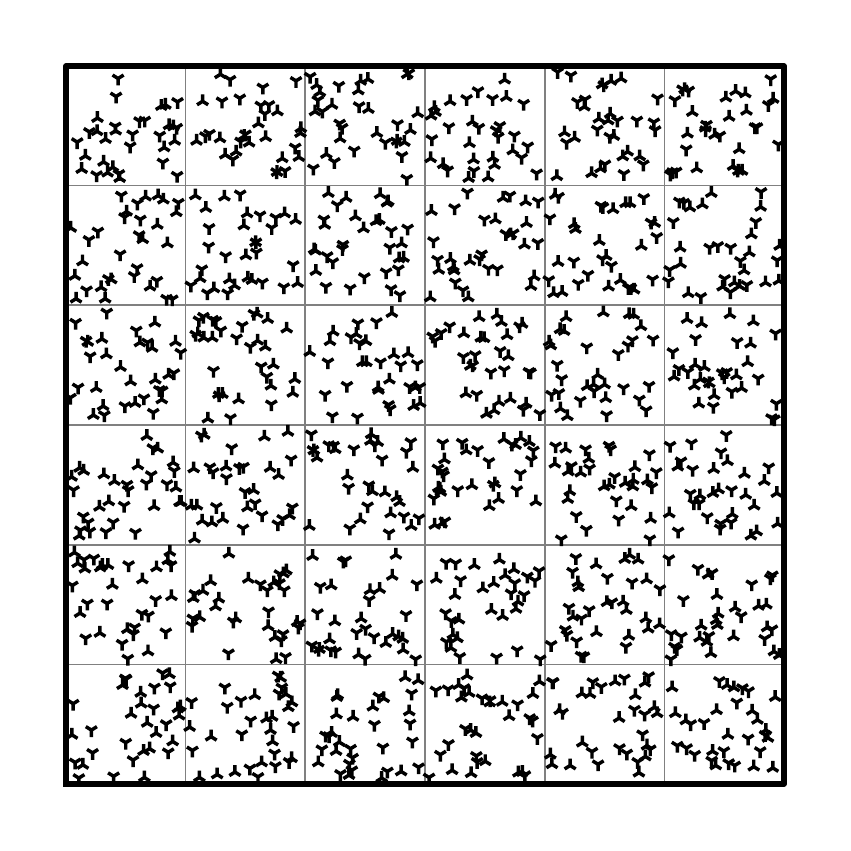}}
\subfloat[600 unit cells $=153.2\,\tt{nm}$]{\includegraphics[width=0.32\widefigwidth]{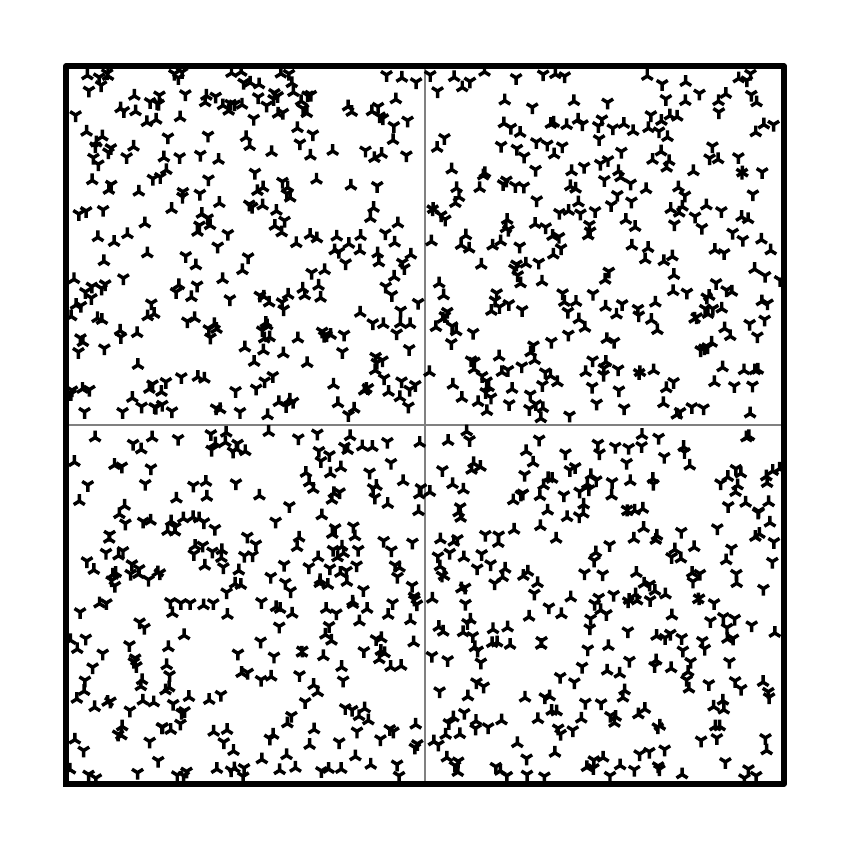}}
\caption{Examples of the initially generated restrictedly-random distributions of dislocations for three sizes of the cells of random distributions, indicated by grey lines. The cell sizes are given below each figure in units of the rotated body-centred tetragonal unit cell, of side length $2.5525\,\tt{\AA}$ and in nm. The two different symbols indicate opposite senses of the burgers vectors, which are $\pm 1/2[110]$ aligned horizontally. The dislocation lines are along $[001]$, into the page.}
\label{fig:restrictedlyrandomdistributions}
\end{center}
\end{figure*}

We now consider our simulations of restrictedly random distributions of dislocations in order to shed light on the meaning of the outer cut-off radius $R_{\textrm{e}}$.
Figure~\ref{fig:restrictedlyrandomdistributions} shows examples of three of our artificially produced restrictedly-random distributions. In the case of the smallest cells (Fig.~\ref{fig:restrictedlyrandomdistributions}(a)) each cell contains a single dislocation dipole and the overall dislocation distribution is highly correlated and homogeneous. As the cell size increases, the distribution becomes less homogeneous and by Fig.~\ref{fig:restrictedlyrandomdistributions}(c) is approaching completely random across the whole supercell.

\begin{figure}\begin{center}
{\includegraphics[width=0.75\figwidth]{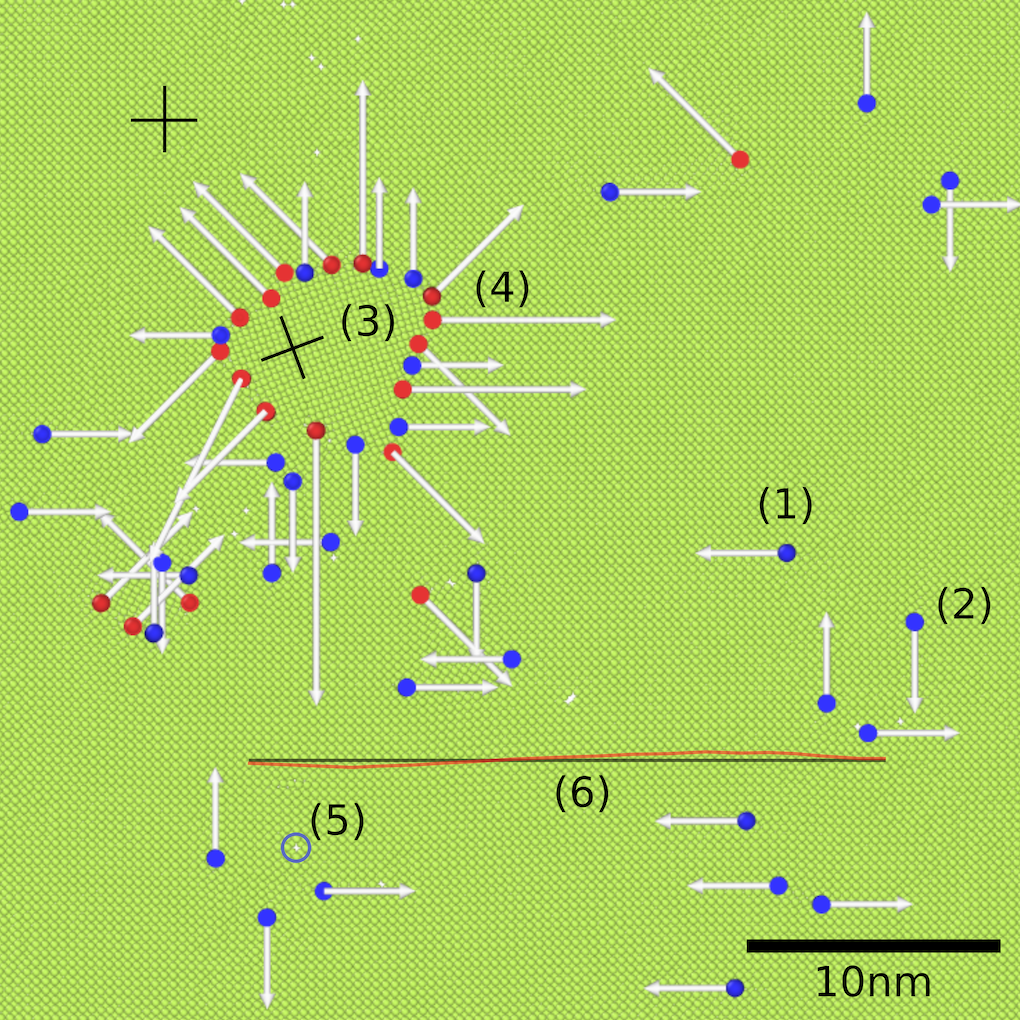}}
\caption{Examples of some of the artefacts that form on relaxation of the restrictedly-random dislocation distributions. Green dots are atoms, and dislocation centres are indicated by blue ($1/2\langle 110 \rangle$) and red ($\langle 100 \rangle$) dots, with white arrows showing their burgers vectors. The numbered features are: (1) a $1/2[ 110 ]$ dislocation of the type inserted by construction; (2) a similar dislocation in a different orientation that has formed by dislocation reaction; (3) a sub-grain with a different orientation to the matrix (crystal orientation indicated by black crosses); (4) Alternative dislocations making up the sub-grain boundary; (5) A vacancy defect; (6) a red line following a $(110)$ lattice plane as an indicator of the distortion in the crystal compared with the straight black line.}
\label{fig:randomissues_150}
\end{center}
\end{figure}

Even with careful optimisation of the annealing time, it is impossible to fully preserve the statistical character of the original dislocation distribution. The dislocations interact strongly and can move significant distances, in some cases aligning to form obvious sub-grain boundaries. These, along with other features of the relaxed distributions, are shown in Figure~\ref{fig:randomissues_150}. In the case of the supercell divided into $2\times 2$ dislocation cells, the large dislocation content gives rise to small ribbon-shaped voids on relaxation (see Figure~\ref{fig:randomissues_600}). As we will see below, these give rise to anomalous diffraction behaviour for this sample. For each sample we use the DXA analysis \cite{0965-0393-18-2-025016} in Ovito to \cite{0965-0393-18-1-015012} to determine the true dislocation density after the relaxation and short anneal. The dislocation density increases somewhat in all cases but remains dominated ($\sim 90\textrm{\%}$) by the $1/2[110]$ perfect dislocations inserted during construction.

\begin{figure}\begin{center}
{\includegraphics[width=0.75\figwidth]{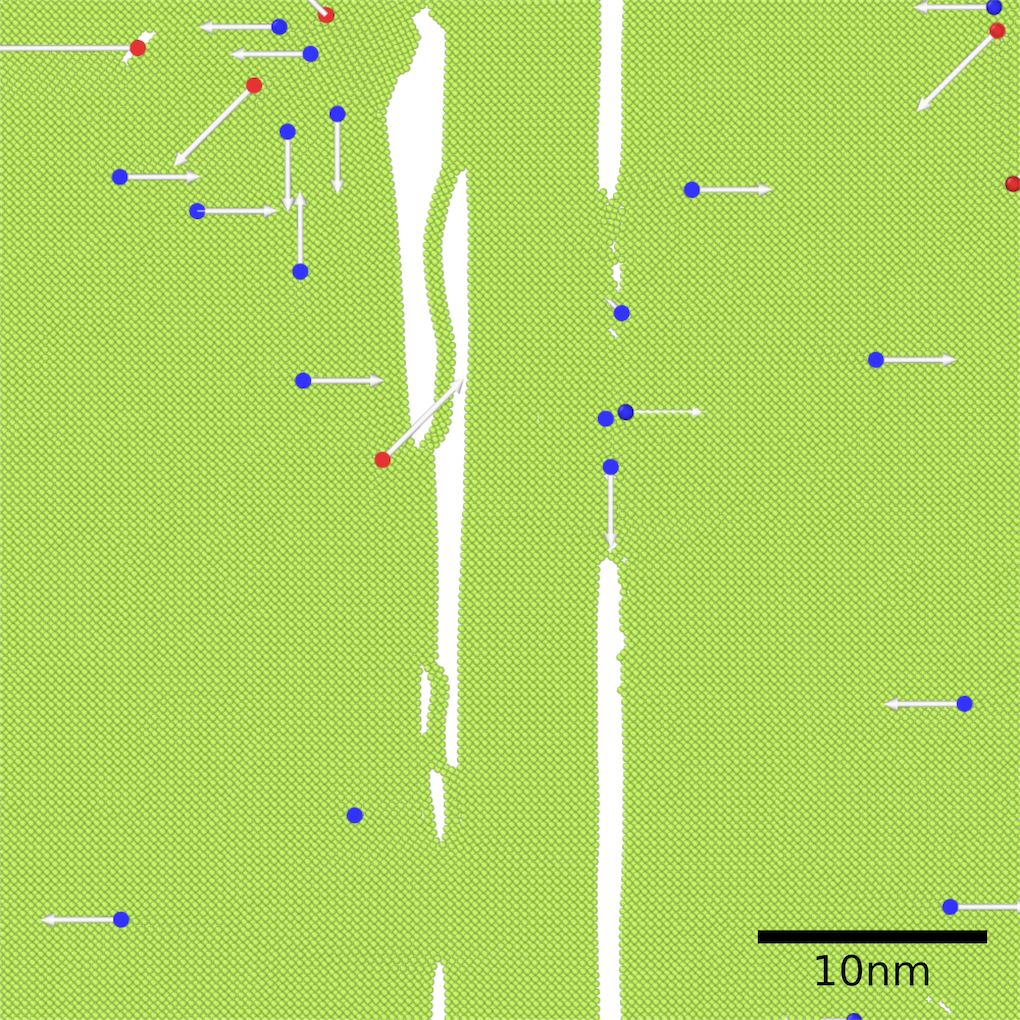}}
\caption{Voids formed during relaxation of the as-constructed restrictedly-random dislocation distribution for the 600 unit cell ($153\,\tt{nm}$) case. Symbols associated with other features are as in Figure~\ref{fig:randomissues_150}.}
\label{fig:randomissues_600}
\end{center}
\end{figure}

\subsubsection{Diffraction peaks}
Because our simulations of restrictedly-random dislocation distributions are carried out in quasi-two-dimensional thin slabs, the real space method produces diffraction line profiles that are dominated by size broadening due to the short dimension. As we have constructed the simulations to contain dislocations of only one burgers vector ($1/2[110]$) we expect the dominant effect of strain broadening to appear in the $(220)$ peak and so we have applied the reciprocal-space approach to calculate the three-dimensional diffraction spot for this peak. Determining the shape of the corresponding line profile peak requires sampling at a high resolution in reciprocal space and we have used a square grid of points, centred on the Bragg peak, with spacing of $10^{-3}\,\tt{\AA}^{-1}$. 

\begin{figure*}\begin{center}
\subfloat[50 unit cells $=12.6\,\tt{nm}$]{\includegraphics[width=0.32\widefigwidth]{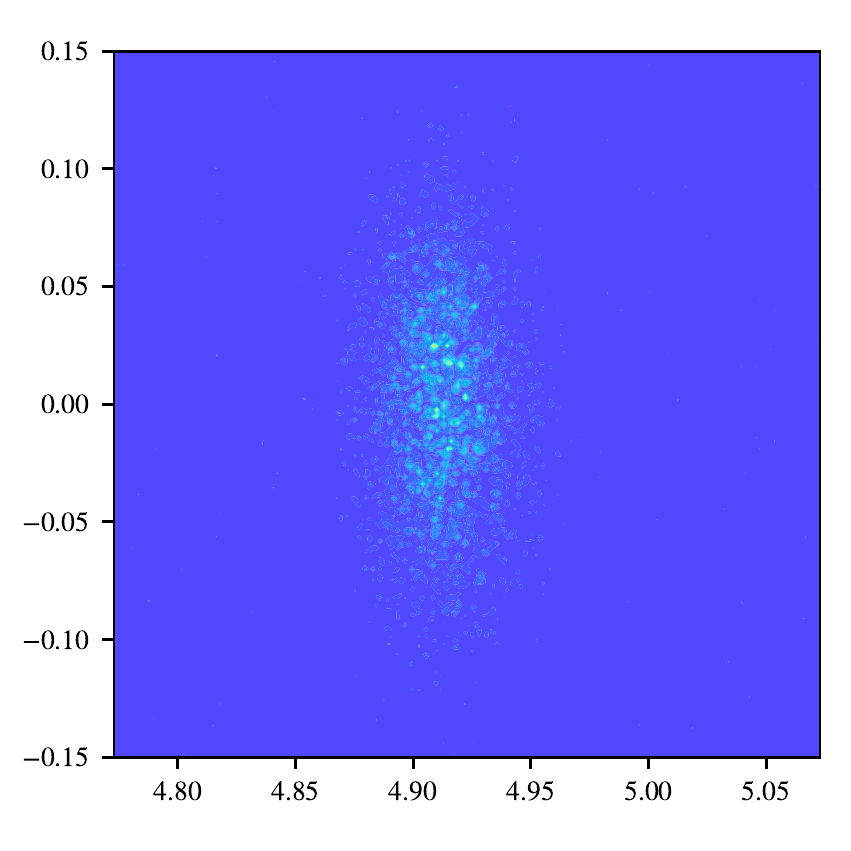}}
\subfloat[200 unit cells $=51.1\,\tt{nm}$]{\includegraphics[width=0.32\widefigwidth]{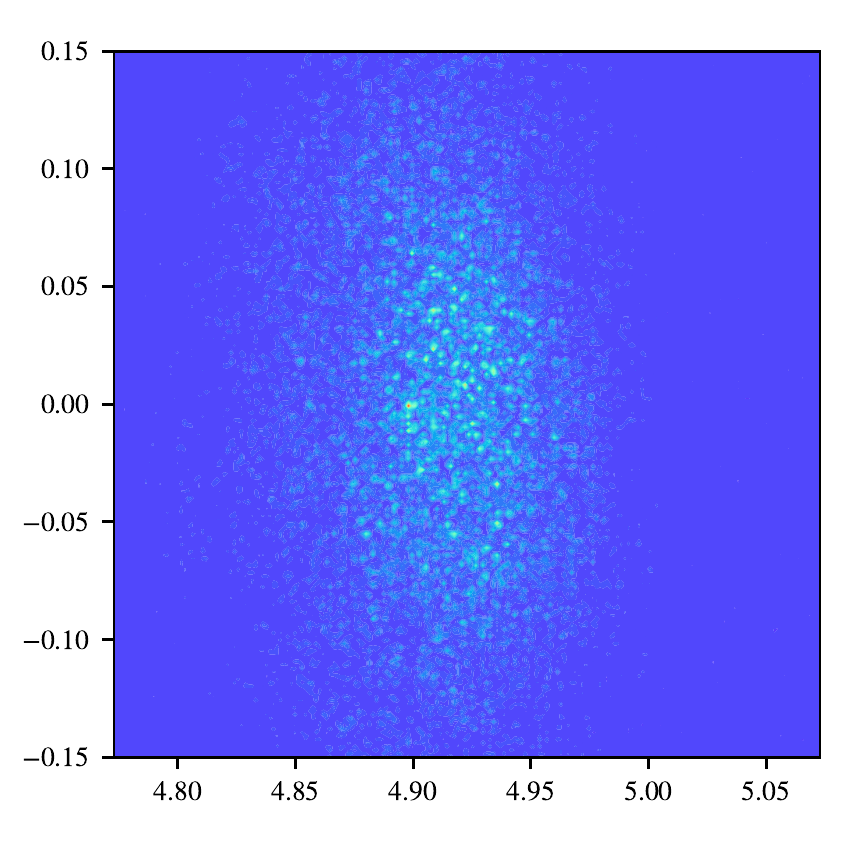}}
\subfloat[600 unit cells $=153.2\,\tt{nm}$]{\includegraphics[width=0.32\widefigwidth]{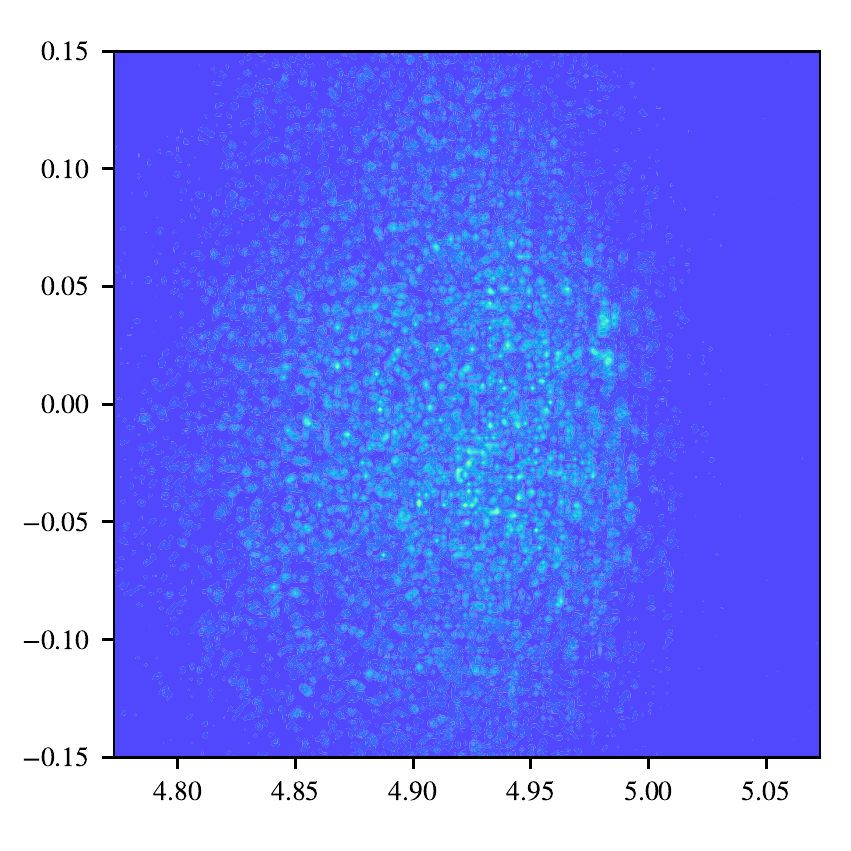}}\hfill
\subfloat[50 unit cells $=12.6\,\tt{nm}$]{\includegraphics[width=0.32\widefigwidth]{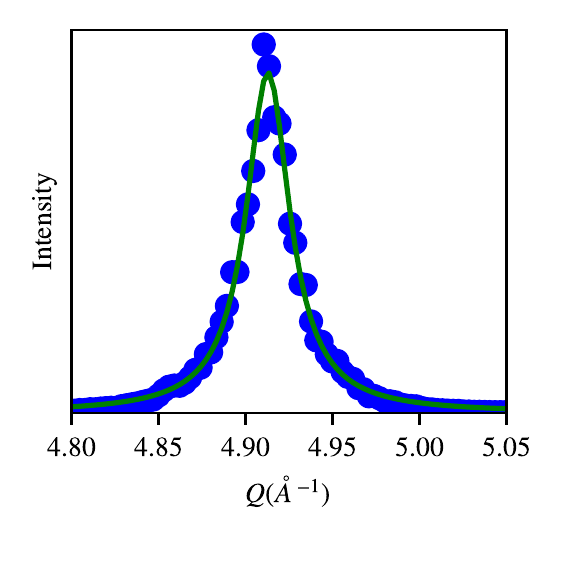}}
\subfloat[200 unit cells $=51.1\,\tt{nm}$]{\includegraphics[width=0.32\widefigwidth]{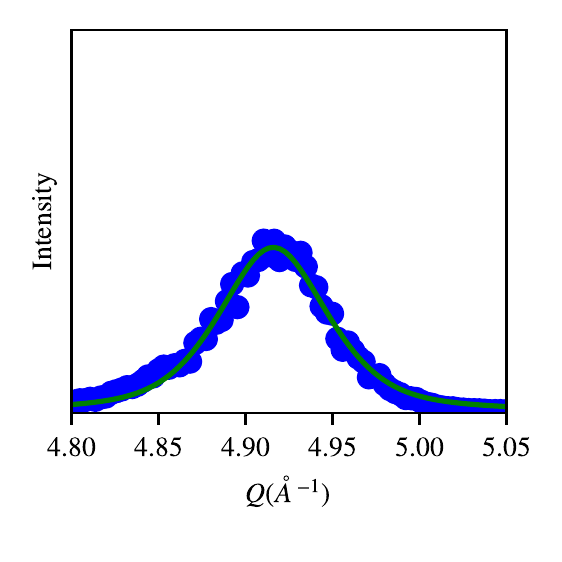}}
\subfloat[600 unit cells $=153.2\,\tt{nm}$]{\includegraphics[width=0.32\widefigwidth]{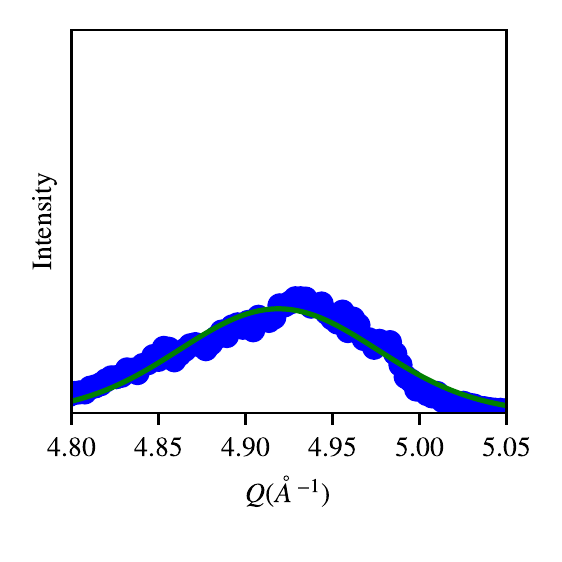}}
\caption{Diffraction patterns for the $(220)$ reflections for three examples of the restrictedly random distributions. (a-c) show projections parallel to the short dimension of the simulation slab of the intensity of the diffraction spots. (d-f) show a reduction of these spots to peaks in the line profile. The green curves are best fits of pseudovoigt functions as a guide to the eye.}
\label{fig:randomdiffraction}
\end{center}
\end{figure*}

Figures~\ref{fig:randomdiffraction}(a-c) shows the projection of the spot into the plane of the slab for the case of three different sizes for the randomly distributed dislocation cells. Figures~\ref{fig:randomdiffraction}(d-f) show the corresponding line profile peaks calculated from the spot data. We see significant variation in the shape of the peaks as the character of the dislocation distribution is varied. Note also that the case of distribution cells of size 600 unit cells ($153.2\,\tt{nm}$) shows an unusual asymmetry that is also visible in the two-dimensional pattern (Figures~\ref{fig:randomdiffraction}(c) and (f)) and is, we assume, the result of the voids present in this sample.

\subsubsection{The outer cut-off radius $R_{\textrm{e}}$}\label{sec:results:const-random:m}

In Section~\ref{sec:methods:lpa:cmwp} we gave the expected form for the Fourier coefficients of the diffraction peaks in \eqref{eqn:logAsimple}, which we repeat here for convenience:
\begin{align}
\ln A_{hkl}(L) &= \ln(1-\alpha L) - \beta \frac{\rho}{\rho_0} L^2 \ln\left( \frac{R_{\tt{e}}}{L} \right), \nonumber
\\
\alpha &= 1/D, \nonumber
\\
\beta &= \frac{G^2_{hkl}\rho_0 C b^2}{8\pi}. \tag{\ref{eqn:logAsimple}}
\end{align}
We can test the validity of this expression by applying it to the diffraction peaks for our restrictedly-random distributions of dislocations. 

\begin{figure}\begin{center}
{\includegraphics[width=\figwidth]{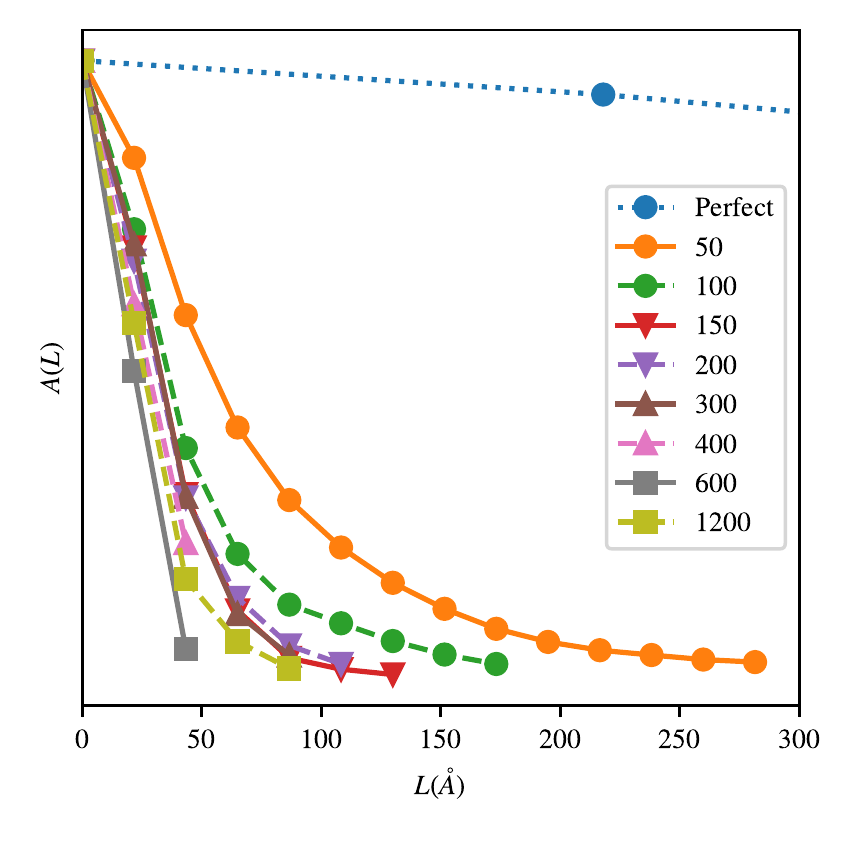}}
\caption{The Fourier coefficients (points) for the $(220)$ peaks for our restrictedly-random dislocation distributions. Lines are a guide for the eye. }
\label{fig:fftprofiles}
\end{center}
\end{figure}

We begin by taking the discrete Fourier transform of the diffraction peaks. The resulting Fourier coefficients are shown in Figure~\ref{fig:fftprofiles}. We note immediately that the 600 unit cell case is once again an anomaly. We can now proceed to fit values for the coefficients $\alpha$ and $\beta$ and for $R_{\textrm{e}}$ in \eqref{eqn:logAsimple}. The value of $\alpha$ depends only on the size broadening and should be the same for all cases. It is easily extracted by a linear fit to the data for the perfect crystal. We find $\alpha = 3.242\times 10^{-4}\,\textrm{\AA}^{-1}$ and we first fix this value. We note that the value expected from theory of $\alpha = 1/D$, where $D$ is the size of the crystal, is $\alpha = 3.265\times 10^{-4}\,\textrm{\AA}^{-1}$, extremely close to the value from a fit to the data.

$\beta$ depends on the burgers vector, the reciprocal lattice vector and contrast factor of the peak under study, and the reference dislocation density. In this case we set $\rho_0$ to the density of dislocations added to the cell \emph{by construction} and use the relative dislocation density \emph{measured} in the cell as an input to \eqref{eqn:logAsimple}. Again, the value of $\beta$ should not vary between the simulations, and to find it we fit $\beta$ and $R_{\tt{e}}$ independently for all eight curves in Fig.~\ref{fig:fftprofiles}, calculating a universal value for $\beta$ as an average across seven of these fits (excluding the 600 unit cell case). We find $\beta = 2.048\times 10^{-4}\,\textrm{\AA}^{-2}$. 

We can compare this to the theoretical value of $\beta$ in \eqref{eqn:logAsimple} in which, for a lattice parameter of $a = 3.61\,\textrm{\AA}$ we have $G_{220} = \sqrt{8}(2\pi/a) = 4.923\,\textrm{\AA}^{-1}$, $b=|(1/2)[110]| = 2.553\,\textrm{\AA}$ and $\rho_0 = 1.228\times 10^{16}\,\textrm{m}^{-2}$. Wilkens \cite{Wilkens:1970aa} provides an estimate for the contrast factor in the case where the $\vec{b}$, $\vec{G}$ and the dislocation line are mutually perpendicular as
\eq
C = \frac{1-4\nu + 8\nu^2}{8(1-\nu)^2},
\qe
where $\nu$ is the Poisson ratio which we calculate from the elastic constants for our model of Cu given by Ackland et al. \cite{Ackland:1987hs}. We find $C=0.117$, which gives a value of $\beta = 0.9032\times 10^{-4}\,\textrm{\AA}^{-2}$. This is within a factor of $\sim 2$ of the value given by a fit to the data; rather good agreement given the approximations involved in deriving \eqref{eqn:logAsimple} and the values of $C$ in the expression for $\beta$.

\begin{figure}\begin{center}
{\includegraphics[width=\figwidth]{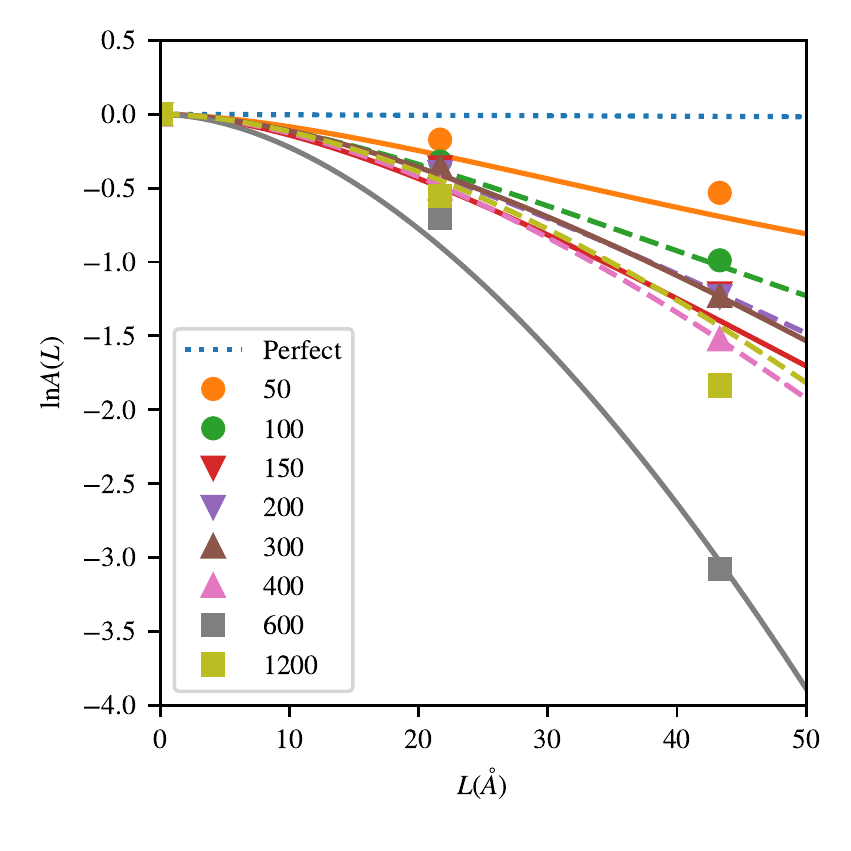}}
\caption{The Fourier coefficients (points) at small $L$ along with best fits (lines) of \eqref{eqn:logAsimple} according to the procedure detailed in the text.}
\label{fig:fftprofilefits}
\end{center}
\end{figure}

Finally, having fixed both $\alpha$ and $\beta$, we refit a value of $R_{\tt{e}}$ for each curve in the low $L$ limit. These final fits are shown in Figure~\ref{fig:fftprofilefits} and are good.

\begin{figure}\begin{center}
{\includegraphics[width=\figwidth]{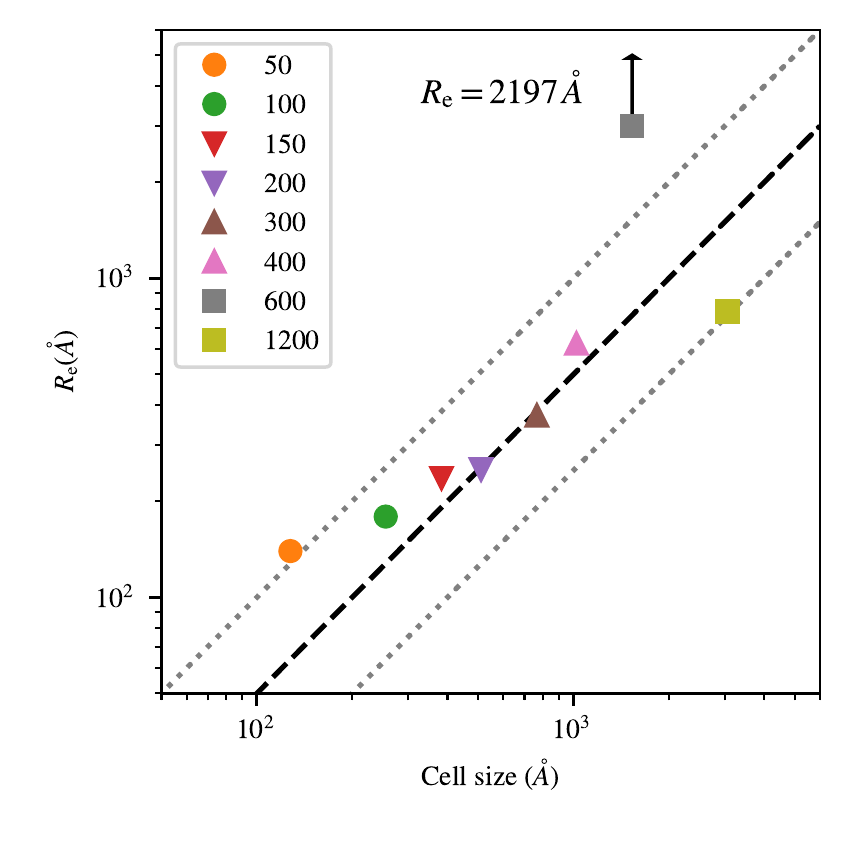}}
\caption{Comparison of the best-fit values for the $R_{\tt{e}}$ parameter in \eqref{eqn:logAsimple} to the cell sizes in the as-constructed restrictedly-random distributions. The black dashed line marks $R_{\mathrm{e}} = 0.5 \times \textrm{cell size}$ and the grey dotted lines are factors of two from this agreement. The point for the anomalous $600$ unit cell case lies far beyond the upper bound of the chart.}
\label{fig:refits}
\end{center}
\end{figure}

In Figure~\ref{fig:refits} we compare the results for the fitting of $R_{\tt{e}}$ with the sizes of the cells used to construct the restrictedly-random dislocation distributions. The correspondence is excellent (except for the anomalous $600$ unit cell case) and provides strong validation for the analytical formulae given in Equations \eqref{eq:wa14}, \eqref{eqn:logAfull}, \eqref{eqn:msqstrain} and \eqref{eqn:logAsimple}. Though in Figure~\ref{fig:refits} we have marked a line corresponding to $R_{\tt{e}} = (1/2)\times\textrm{cell size}$ as a reference for agreement, we note that there is a difference between the infinite cylindrical regions invoked by Wilkens in discussion of $R_{\tt{e}}$ \cite{Wilkens:1970aa} and the effectively infinitely long cuboidal cells used in our simulations. Furthermore, expecting perfect correspondence would be ambitious, given the approximations made in deriving \eqref{eqn:logAsimple} and the relaxations that take place in our simulations and shift the dislocation distribution away from the as-constructed form.

\section{Conclusions}\label{sec:conclusions}
We have built atomistic models of copper crystals containing cold-work defects and restrictedly-random dislocation distributions. By generating x-ray diffraction profiles for these model crystals and applying experimental procedures to their analysis we have been able to compare the inferences from line profile analysis to the ``ground truth'' of defect content measured directly from the atomistic structure. We have also been able to test interpretations of the outer cut-off radius $R_{\tt{e}}$ (or, equivalently, Wilkens' $M$ parameter) and the analytical expressions for their effect on line-profile peak shape. We found:
\begin{enumerate}
\item that the dislocation density for cold-worked material derived from CMWP line profile analysis agreed well with the known density across two orders of magnitude;
\item that the stacking fault density for cold-worked material derived from CMWP line profile analysis agreed very well with the known density across two orders of magnitude;
\item that interpretation of the meaning of $R_{\tt{e}}$ (or the $M$ parameter) in terms of dipole character in a restrictedly-random dislocation distribution and its incorporation into analytical models for the Fourier coefficients for peak shape change was directly validated by our models.
\end{enumerate}

Overall, our results represent an important verification of the ability of CMWP to quantitatively assess the concentrations of line (dislocations) and planar (stacking faults) in crystalline materials. More generally, our work demonstrates a productive approach to validating and benchmarking a powerful experimental analysis technique, which could be successfully applied to a broader class of defects and in other materials.

\begin{acknowledgments}
C.P.R. was funded by a University Research Fellowship of The Royal Society. T.U. was funded by the EPSRC Programme Grant (MIDAS: EP/S01702X/1). G.R. is grateful for the support of OTKA grant K124926 funded by the Hungarian National Research, Development and Innovation Office (NKFIH). Simulations were carried out on the University of Manchester's Computational Shared Facility. The core research data, along with computer scripts implementing our methods, are freely available for download \cite{Race:tb}.
\end{acknowledgments}

%\section{Appendix}

%\section*{References}
\bibliographystyle{unsrt}
\bibliography{references}

\begin{thebibliography}{10}

\bibitem{Wilkens:1970aa}
M.~Wilkens.
\newblock The determination of density and distribution of dislocations in
  deformed single crystals from broadened x-ray diffraction profiles.
\newblock {\em physica status solidi (a)}, 2(2):359--370, 2020/01/06 1970.

\bibitem{Bacon:1980fk}
D.~J. Bacon, D.~M. Barnett, and R.~O. Scattergood.
\newblock Anisotropic continuum theory of lattice defects.
\newblock {\em Progress in Materials Science}, 23:51--262, 1980.

\bibitem{Krivoglaz:1963aa}
M.A. Krivoglaz and K.P. Rjaboshapka.
\newblock Theory of x-ray scattering by crystals containing dislocations, screw
  and edge dislocations randomly distributed throughout the crystal.
\newblock {\em Fiz. Metallov Metalloved.}, 15:18--31, 1963.

\bibitem{Wilkens:1968aa}
M~Wilkens and M.~O Bargouth.
\newblock Die bestimmung der versetzungsdichte verformter kupfereinkristalle
  aus verbreiterten r{\"o}ntgenbeugungsprofilen.
\newblock {\em Acta Metallurgica}, 16(3):465--468, 1968.

\bibitem{Wilkens:1969ab}
M~Wilkens.
\newblock The mean square stresses for a completely random and a restrictedly
  random distribution of dislocations in a cylindrical body.
\newblock In J~A Simmons, R~deWit, and R~Bullough, editors, {\em Fundamental
  aspects of dislocation theory}, volume~II, pages 1191--1193, 1969.

\bibitem{Wilkens:1969ac}
M~Wilkens.
\newblock Theoretical aspects of kinematical x-ray diffraction profiles from
  crystals containing dislocation distributions.
\newblock In J~A Simmons, R~deWit, and R~Bullough, editors, {\em Fundamental
  aspects of dislocation theory}, volume~II, pages 1195--1221, 1969.

\bibitem{Essmann:1966vd}
U.~Essmann.
\newblock Elektronenmikroskopische untersuchung der versetzungsanordnung
  verformter kupferein kristalle iii. bestimmung der versetzungsdichte.
\newblock {\em physica status solidi (b)}, 17(2):725--737, 2022/03/15 1966.

\bibitem{Essmann:1965wi}
U.~Essmann.
\newblock Elektronenmikroskopische untersuchung der versetzungsanordnung
  verformter kupfereinkristalle ii. die versetzungsanordnung im bereich ii.
\newblock {\em physica status solidi (b)}, 12(2):723--747, 2022/03/15 1965.

\bibitem{Gottler:1973uh}
E.~G{\"o}ttler.
\newblock Versetzungsstruktur und verfestigung von
  {$[$}100{$]$}-kupfereinkristallen.
\newblock {\em The Philosophical Magazine: A Journal of Theoretical
  Experimental and Applied Physics}, 28(5):1057--1076, 11 1973.

\bibitem{Ungar:1984aa}
T.~Ungar, H.~Mughrabi, D.~R{\"o}nnpagel, and M.~Wilkens.
\newblock X-ray line-broadening study of the dislocation cell structure in
  deformed {$[$}001{$]$}-orientated copper single crystals.
\newblock {\em Acta Metallurgica}, 32(3):333--342, 1984.

\bibitem{0965-0393-18-1-015012}
Alexander Stukowski.
\newblock Visualization and analysis of atomistic simulation data with
  ovito--the open visualization tool.
\newblock {\em Modelling and Simulation in Materials Science and Engineering},
  18(1):015012, 2010.

\bibitem{0965-0393-18-2-025016}
Alexander Stukowski and Karsten Albe.
\newblock Dislocation detection algorithm for atomistic simulations.
\newblock {\em Modelling and Simulation in Materials Science and Engineering},
  18(2):025016, 2010.

\bibitem{Frenkel:2002aa}
Daan Frenkel and Berend Smit.
\newblock {\em {Understanding molecular simulation from algorithms to
  applications}}, pages 2--3.
\newblock Academic Press, San Diego, 2002.

\bibitem{Note1}
An understandable trend, given the increasing ease and decreasing cost with
  which ``ab initio'' type calculations can now be performed.

\bibitem{Zhang:2020aa}
Zhenbo Zhang, {\'E}va {\'O}dor, Diana Farkas, Bertalan J{\'o}ni, G{\'a}bor
  Rib{\'a}rik, G{\'e}za Tichy, Sree-Harsha Nandam, Julia Ivanisenko, Michael
  Preuss, and Tam{\'a}s Ung{\'a}r.
\newblock Dislocations in grain boundary regions: The origin of heterogeneous
  microstrains in nanocrystalline materials.
\newblock {\em Metallurgical and Materials Transactions A}, 51(1):513--530,
  2020.

\bibitem{Kamminga:2000aa}
J.-D. Kamminga and R.~Delhez.
\newblock Calculation of diffraction line profiles from specimens with
  dislocations. a comparison of analytical models with computer simulations.
\newblock {\em Journal of Applied Crystallography}, 33(4):1122--1127, 2000.

\bibitem{Nabarro:1967aa}
F.~R.~N. Nabarro.
\newblock {\em Theory of Crystal Dislocations}.
\newblock Oxford: Clarendon Press, 1967.

\bibitem{Ackland:1987hs}
G.~J. Ackland, G.~Tichy, V.~Vitek, and M.~W. Finnis.
\newblock Simple n-body potentials for the noble metals and nickel.
\newblock {\em Philosophical Magazine A}, 56(6):735--756, 2013/01/25 1987.

\bibitem{Groma:1988aa}
I.~Groma, T.~Ungar, and M.~Wilkens.
\newblock Asymmetric x-ray line broadening of plastically deformed crystals. i.
  theory.
\newblock {\em Journal of Applied Crystallography}, 21(1):47--54, 1988.

\bibitem{Ungar:1989aa}
T.~Ung{\'a}r, I.~Groma, and M.~Wilkens.
\newblock Asymmetric x-ray line broadening of plastically deformed crystals.
  ii. evaluation procedure and application to {$[$}001{$]$}-cu crystals.
\newblock {\em Journal of Applied Crystallography}, 22(1):26--34, 2020/05/26
  1989.

\bibitem{Balogh:2006aa}
Levente Balogh, G{\'a}bor Rib{\'a}rik, and Tam{\'a}s Ung{\'a}r.
\newblock Stacking faults and twin boundaries in fcc crystals determined by
  x-ray diffraction profile analysis.
\newblock {\em Journal of Applied Physics}, 100(2):023512, 2020/05/11 2006.

\bibitem{Plimpton:1995fv}
Steve Plimpton.
\newblock Fast parallel algorithms for short-range molecular dynamics.
\newblock {\em Journal of Computational Physics}, 117(1):1--19, 3 1995.

\bibitem{Race:2019ac}
C~P Race.
\newblock Atomistic simulations of grain boundary migration under
  recrystallisation conditions.
\newblock {\em Modelling and Simulation in Materials Science and Engineering},
  27(6):064002, 2019.

\bibitem{Debye:1915aa}
P.~Debye.
\newblock Zerstreuung von r{\"o}ntgenstrahlen.
\newblock {\em Annalen der Physik}, 351(6):809--823, 2021/02/04 1915.

\bibitem{Race:tb}
C~P Race.
\newblock https://doi.org/10.5281/zenodo.6362428.

\bibitem{Warren:1952aa}
B.~E. Warren and B.~L. Averbach.
\newblock The separation of cold‐work distortion and particle size broadening
  in x‐ray patterns.
\newblock {\em Journal of Applied Physics}, 23(4):497--497, 2020/05/13 1952.

\bibitem{Williamson:1953aa}
G.~K Williamson and W.~H Hall.
\newblock X-ray line broadening from filed aluminium and wolfram.
\newblock {\em Acta Metallurgica}, 1(1):22--31, 1953.

\bibitem{Warren:1950aa}
B.~E. Warren and B.~L. Averbach.
\newblock The effect of cold‐work distortion on x‐ray patterns.
\newblock {\em Journal of Applied Physics}, 21(6):595--599, 2020/05/14 1950.

\bibitem{Warren:1959aa}
B.~E. Warren.
\newblock X-ray studies of deformed metals.
\newblock {\em Progress in Metal Physics}, 8:147--202, 1959.

\bibitem{Delhez:1976wh}
R.~Delhez and E.~J. Mittemeijer.
\newblock The elimination of an approximation in the warren-averbach analysis.
\newblock {\em Journal of Applied Crystallography}, 9(3):233--234, 1976.

\bibitem{Berkum:1994ts}
J.~G.~M. van Berkum, A.~C. Vermeulen, R.~Delhez, T.~H. de~Keijser, and E.~J.
  Mittemeijer.
\newblock Applicabilities of the warren-averbach analysis and an alternative
  analysis for separation of size and strain broadening.
\newblock {\em Journal of Applied Crystallography}, 27(3):345--357, 1994.

\bibitem{Balzar:2004tg}
D.~Balzar, N.~Audebrand, M.~R. Daymond, A.~Fitch, A.~Hewat, J.~I. Langford,
  A.~Le~Bail, D.~Louer, O.~Masson, C.~N. McCowan, N.~C. Popa, P.~W. Stephens,
  and B.~H. Toby.
\newblock Size-strain line-broadening analysis of the ceria round-robin sample.
\newblock {\em Journal of Applied Crystallography}, 37(6):911--924, 2004.

\bibitem{Ungar:2001aa}
T.~Ungar, J.~Gubicza, G.~Ribarik, and A.~Borbely.
\newblock Crystallite size distribution and dislocation structure determined by
  diffraction profile analysis: principles and practical application to cubic
  and hexagonal crystals.
\newblock {\em J Appl. Cryst}, 34(3):298--310, 2001.

\bibitem{Ungar:1999aa}
T.~Ungar, I.~Dragomir, A.~Revesz, and A.~Borbely.
\newblock The contrast factors of dislocations in cubic crystals: the
  dislocation model of strain anisotropy in practice.
\newblock {\em Journal of Applied Crystallography}, 32(5):992--1002, 1999.

\bibitem{Ribarik:2020aa}
G{\'a}bor Rib{\'a}rik, Bertalan J{\'o}ni, and Tam{\'a}s Ung{\'a}r.
\newblock The convolutional multiple whole profile (cmwp) fitting method, a
  global optimization procedure for microstructure determination.
\newblock {\em Crystals}, 10(7), 2020.

\bibitem{Krivoglaz:1996ug}
Mikhail~A. Krivoglaz.
\newblock {\em X-Ray and Neutron Diffraction in Nonideal Crystals}.
\newblock Springer, Berlin, Heidelberg, 1996.

\bibitem{Groma:1997tk}
I.~Groma.
\newblock Link between the microscopic and mesoscopic length-scale description
  of the collective behavior of dislocations.
\newblock {\em Physical Review B}, 56(10):5807--5813, 09 1997.

\bibitem{Groma:2016uc}
Istv{\'a}n Groma, Michael Zaiser, and P{\'e}ter~Dus{\'a}n Isp{\'a}novity.
\newblock Dislocation patterning in a two-dimensional continuum theory of
  dislocations.
\newblock {\em Physical Review B}, 93(21):214110--, 06 2016.

\bibitem{Zaiser:2001ur}
M.~Zaiser, M.~Carmen Miguel, and I.~Groma.
\newblock Statistical dynamics of dislocation systems: The influence of
  dislocation-dislocation correlations.
\newblock {\em Physical Review B}, 64(22):224102--, 11 2001.

\bibitem{Zepeda-Ruiz:2020aa}
Luis~A. Zepeda-Ruiz, Alexander Stukowski, Tomas Oppelstrup, Nicolas Bertin,
  Nathan~R. Barton, Rodrigo Freitas, and Vasily~V. Bulatov.
\newblock Atomistic insights into metal hardening.
\newblock {\em Nature Materials}, 2020.

\bibitem{Ungar:1996aa}
T.~Ung{\'a}r and A.~Borb{\'e}ly.
\newblock The effect of dislocation contrast on x‐ray line broadening: A new
  approach to line profile analysis.
\newblock {\em Applied Physics Letters}, 69(21):3173--3175, 2020/06/16 1996.

\end{thebibliography}
%
%\begin{thebibliography}{10}
%\end{thebibliography}

\end{document}